\documentclass[12pt]{article}

\usepackage{graphics}
\usepackage{cite} 
\newenvironment{Eqnarray}%
     {\arraycolsep 0.14em\begin{eqnarray}}{\end{eqnarray}}

\makeatletter
\@addtoreset{equation}{section}

\makeatother

\begin{document}

\def\nicefrac#1#2{\hbox{${#1\over #2}$}}

\setlength{\textwidth}{16.8 cm}
\setlength{\textheight}{23 cm}
\addtolength{\topmargin}{-2.5 cm}

\newcommand{\nn}{\nonumber}
\newcommand{\raw}{\rightarrow}
\newcommand{\be}{\begin{equation}}
\newcommand{\ee}{\end{equation}}
\newcommand{\bea}{\begin{Eqnarray}}
\newcommand{\eea}{\end{Eqnarray}}
\newcommand{\dl}{\stackrel{\leftarrow}{D}}
\newcommand{\dr}{\stackrel{\rightarrow}{D}}
\newcommand{\dd}{\displaystyle}
\def\lsim{\mathrel{\raise.3ex\hbox{$<$\kern-.75em\lower1ex\hbox{$\sim$}}}}
\def\gsim{\mathrel{\raise.3ex\hbox{$>$\kern-.75em\lower1ex\hbox{$\sim$}}}}
\def\ifmath#1{\relax\ifmmode #1\else $#1$\fi}
\def\ls#1{\ifmath{_{\lower1.5pt\hbox{$\scriptstyle #1$}}}}

\pagestyle{empty}
\begin{flushright}
FERMILAB-Pub-00/138-T  \\
FTUAM 99-27 \\
IFT-UAM 00-20  \\
SCIPP 00/22     \\
hep--ph/0007006 \\
July 2000 \\
\end{flushright}
\vskip1cm

\renewcommand{\thefootnote}{\fnsymbol{footnote}}
\begin{center}
{\Large\bf
SUSY-QCD corrections to the MSSM \boldmath$h^0 b \bar b$ vertex
in the decoupling limit}\\[1cm]
{\large Howard E. Haber$^a$, Mar\'{\i}a J. Herrero$^b$, 
Heather E. Logan$^c$,
\\[6pt]
Siannah Pe\~naranda$^b$, Stefano Rigolin$^d$
and David Temes$^b$~\footnote{electronic addresses:
haber@scipp.ucsc.edu, herrero@delta.ft.uam.es,
logan@fnal.gov, siannah@delta.ft.uam.es, srigolin@umich.edu,
temes@delta.ft.uam.es}
}\\[6pt]
{\it $^a$ Santa Cruz Institute for Particle Physics  \\
   University of California, Santa Cruz, CA 95064, USA. \\
$^b$ Departamento de F\'{\i}sica Te\'{o}rica \\
   Universidad Aut\'{o}noma de Madrid,
   Cantoblanco, 28049 Madrid, Spain. \\
$^c$ Theoretical Physics Department \\
   Fermi National Accelerator Laboratory, Batavia, IL 60510-0500, USA.\\
$^d$ Department of Physics \\
   University of Michigan, Ann Arbor, MI 48109, USA.}
\\[1cm]

{\bf Abstract}
\end{center}

\enlargethispage*{1000pt}
We analyze the supersymmetric (SUSY) QCD contribution to the $h^0 b
\bar{b}$ coupling at one loop in the Minimal Supersymmetric Model
(MSSM) in the decoupling limit.  Analytic expressions in the large
SUSY mass region are derived and the decoupling behavior of the
corrections is examined in various limiting cases, where some or all
of the SUSY mass parameters become large.  We show that in the
decoupling limit of large SUSY mass parameters and large CP-odd Higgs
mass, the $h^0 b \bar b$ coupling approaches its Standard Model value
at one loop.  However, the onset of decoupling is delayed when
$\tan\beta$ is large.  In addition, the one-loop SUSY-QCD corrections
decouple if the masses of either the bottom squarks or the gluino are
separately taken large; although the approach to decoupling is
significantly slower in the latter case.

\vfill
\clearpage

\renewcommand{\thefootnote}{\arabic{footnote}}
\setcounter{footnote}{0}

\pagestyle{plain}

\section{Introduction}
Once a light CP-even Higgs boson is discovered, it will be crucial to
measure
as many of its couplings as possible with the highest accuracy possible.
By measuring the Higgs couplings to gauge bosons, one can learn
whether only one
Higgs boson is involved in electroweak symmetry breaking dynamics.
Moreover, the Higgs couplings to vector bosons are sensitive to the
possible existence of non-doublet isospin structure in the Higgs
sector.  By measuring the Higgs couplings to fermion pairs, one can
learn whether the Higgs boson
is responsible for fermion mass generation.  Knowledge of the
trilinear and quartic Higgs self-couplings, although extremely
difficult
to obtain, would allow one to reconstruct the Higgs potential and
directly test the mechanism of electroweak symmetry breaking.
Finally, if the couplings can be measured at the level of the radiative
corrections, one could then derive significant constraints on new physics
beyond the reach of the present accelerators.
A detailed study of radiative corrections to the Higgs couplings would be
especially important if a light Higgs boson were discovered in
the mass range predicted by the minimal supersymmetric extension of
the Standard Model (MSSM),
but supersymmetric (SUSY) particles were not found.  In this case, the precise
experimental determination of Higgs couplings could provide
indirect information about the
preferred region of SUSY parameter space.  For example, one could
predict (in the context of the MSSM)
whether the data favored a SUSY spectrum below the 1 TeV energy scale.

It is well known that the tree-level couplings of the lightest
MSSM Higgs boson ($h^0$) to fermion pairs and gauge bosons
tend to their Standard Model (SM) values
in the decoupling limit, $M_A \gg M_Z$ \cite{decoupling}, where
$M_A$ is the mass of the CP-odd neutral Higgs boson ($A^0$) of the MSSM.
As a consequence of this decoupling, distinguishing
the lightest MSSM Higgs boson in the large $M_A$
limit from the Higgs boson of the Standard Model (SM) will be very
difficult.
Formally, the decoupling of all SUSY particles
(including the radiative corrections) implies that
in the effective low-energy theory, all observables tend
to their SM values in the limit of large SUSY masses and large $M_A$.
It has been shown that all of the SUSY particles in the MSSM, including
the heavy Higgs bosons $H^0$, $A^0$ and $H^\pm$, decouple at one-loop order
from the low-energy electroweak gauge boson physics \cite{dhp}.
In particular, the contributions of the SUSY particles
to low-energy processes either fall as inverse powers of the SUSY
mass parameters or can be absorbed into counterterms for the tree-level
couplings of the low-energy theory \cite{ac}.  As a result,
the radiative corrections involving SUSY particles go to zero
in the asymptotic large SUSY mass limit.
Our aim is to determine the nature of the decoupling limit
at one-loop for the couplings of $h^0$ to SM particles.

In this paper, we focus on the
$h^0$ coupling to $b\bar b$.  This coupling determines the partial width
of $h^0 \to b\bar b$, which is by far the dominant decay mode of $h^0$ in most
of the MSSM parameter space.
Because this decay is dominant, accurate knowledge of the $h^0 b \bar b$
coupling is very important for Higgs boson searches.  At LEP and
the Tevatron, the primary Higgs search channel is $h^0 \to b \bar b$.
The experimental reach of Higgs boson searches at the upcoming Tevatron
Run 2 depends critically on the $h^0 \to b\bar b$
branching ratio \cite{cmwpheno}.
In contrast, the Higgs boson searches at LEP do not depend
as critically on the $h^0 \to b \bar b$ decay.  At LEP, there
is sufficient cross-section to detect the Higgs boson in multiple
channels. Moreover, even without observing the Higgs decay products,
the Higgs boson mass can be reconstructed by detecting
the recoiling $Z$ boson in $e^+ e^- \to Z h^0$.
At the CERN LHC, the primary Higgs search channel
in the mass region below 130~GeV is the rare decay
$h^0 \to \gamma \gamma$.  The Higgs event rate in this channel
is affected strongly if the total
width of $h^0$ is modified due to corrections to the dominant $b\bar b$
decay mode \cite{cmwcomplementarity}.  The same holds true for other
search channels at the LHC such as $h^0 \to \tau^+ \tau^-$ \cite{Htotautau}.

In this paper we study the MSSM radiative corrections
to the $h^0 b \bar b$ coupling at one loop, to leading order
in $\alpha_s$, and we analyze in detail their behavior in the 
decoupling limit.\footnote{The radiatively-corrected $h^0 b\bar b$
coupling in the decoupling limit has also been considered previously
in refs.~\cite{cmwpheno,polonsky,kolda}.}
These corrections are due to the SUSY-QCD (SQCD) sector and arise from
gluino and bottom-squark
(sbottom) exchange.  Because of the dependence on the strong
coupling constant, these are expected to be the most significant
one-loop MSSM contributions over much of the MSSM parameter space.
Potentially significant contributions can also arise from the
SUSY-electroweak sector (the most significant of which are proportional
to the Higgs-top quark Yukawa coupling); we will
address these corrections elsewhere and do not consider them here.

The SQCD corrections to the $h^0 b \bar b$ coupling were first
calculated using a diagrammatic approach in ref.~\cite{Dabelstein}
(which also contains results for the SUSY-electroweak corrections).  
These corrections have also been obtained in refs.~\cite{cjs,polonsky}.
The SQCD corrections were also computed in an effective
Lagrangian approach in
ref.~\cite{cmwpheno}, using the SUSY contributions to the $b$--quark
self energy obtained in refs.~\cite{hrs-h-copw,pbmz} and
neglecting terms suppressed by inverse powers of SUSY masses.

The radiatively-corrected $h^0 b \bar b$ coupling depends on the
CP-even Higgs mixing angle $\alpha$.  At tree-level, this mixing
angle is determined
by fixing $\tan\beta$ and $M_A$.  At one-loop order, there are no
$\mathcal{O}(\alpha_s)$ corrections to this mixing angle.
As a result, working to leading order in $\alpha_s$, we may employ
tree-level relations for $\alpha$ in our computation
of the $h^0 b \bar b$ coupling.  This procedure is no longer adequate
once one-loop SUSY-electroweak effects are included.
In the latter case, the one-loop radiative corrections to $\alpha$
must be taken into account,
as described in refs.~\cite{cmwpheno,cmwcomplementarity,hhw}.
These papers show that the interplay between the
radiative corrections to the mixing angle and to the $h^0 b \bar b$
coupling can be very important for Higgs collider phenomenology,
particularly in the case when the branching ratio for $h^0\to b\bar b$
is suppressed.\footnote{The relevance of the suppressed $h^0 b \bar b$
coupling for phenomenology has also been emphasized in 
refs.~\cite{wells1,wells2,kolda}.}  This is most easily seen as
follows.  When radiative corrections to the mixing angle $\alpha$ are
included, it becomes possible to tune this angle to zero independent
of the value of $\tan\beta$ by varying the SUSY parameters.
At $\alpha = 0$, the tree-level couplings
of $h^0$ to $b \bar b$ and $\tau^+ \tau^-$ vanish, as do the
ordinary QCD corrections \cite{qcd} to the $h^0 b \bar b$ coupling.
However, because the SQCD corrections to the $h^0 b \bar b$ coupling
include contributions from diagrams involving the $h^0$ coupling to
sbottoms, they remain nonzero at $\alpha = 0$.
As a result, the $h^0 \tau^+ \tau^-$ coupling goes to zero at a different
point in SUSY parameter space than the $h^0 b \bar b$ coupling does
\cite{hhw,kolda,cmwcomplementarity}.  We will come back to these issues
and study the approach to decoupling of the SUSY-electroweak corrections
in a later paper.


In some regions of the MSSM parameter space, the SQCD corrections
to the $h^0 b \bar b$ coupling become so large that it is important
to take into account higher-order corrections.  This has been carried
out in refs.~\cite{cgnw,ehkmy} by
resumming the leading $\tan\beta$ contributions to all orders of perturbation
theory using an effective Lagrangian approach.  This
resummation is not important in our present work because
we are interested in the decoupling limit in which the one-loop
corrections to the $h^0 b\bar b$ coupling are small.

In this paper we analyze the full diagrammatic formulae for the
on-shell one-loop SQCD corrections to the $h^0 b \bar b$ coupling.  We
perform expansions in inverse powers of SUSY masses in order to
examine the decoupling behavior when the SUSY masses are large
compared to $M_Z$.  The SQCD corrections depend on a number of
different SUSY mass parameters, and the relative sizes of these
masses affect the manifestation of the decoupling.  To remain as
model-independent as possible, we make no assumptions about relations
among the SUSY parameters that may arise from grand unification or
specific SUSY-breaking scenarios.  We consider the soft-SUSY-breaking
parameters and the $\mu$ parameter to be independent parameters whose
magnitudes are all of order 1~TeV.

In this paper, we demonstrate that in the limit of large
$M_A$ (in this limit one also has $M_{H^0},M_{H^\pm} \gg M_Z$)
and large sbottom and gluino masses ($M_{\tilde b_i},M_{\tilde g} \gg M_Z$),
the SM expression for the $h^0b\bar b$ one-loop coupling is recovered.
That is, the SQCD corrections to the $h^0 b \bar b$
coupling decouple in the limit of large SUSY masses and large $M_A$.
In particular, we examine the case of large $\tan\beta$, for which
the SQCD corrections are enhanced.
This enhancement can delay the onset of
decoupling and give rise to a significant one-loop correction, even for
moderate to large values of the SUSY masses.

This paper is organized as follows.
In Section~\ref{sec:masses} we define our notation and briefly review
the Higgs and sbottom sectors of the MSSM.
In Section~\ref{sec:exactcorr} we give the exact one loop formula for
the SQCD corrections to the $h^0 b \bar b$ coupling.
In Section~\ref{sec:analytic} we derive analytic expressions for
the SQCD corrections in the limit of large SUSY masses.  We analyze
the decoupling of the SQCD corrections for various hierarchies of
mass parameters, and numerically compare the analytic approximations to
the exact one-loop result.
In Section~\ref{sec:conclusions} we summarize our conclusions.
Finally, the Appendix contains expansions
of the one-loop integrals used in our calculations.

\section{Higgs and sbottom masses in the MSSM}
\label{sec:masses}

In the MSSM, the parameters of the Higgs sector are constrained at
tree-level in such a way that the Higgs masses and mixing angles
depend on only two unknown parameters.
These are commonly chosen to be the mass
of the CP-odd neutral Higgs boson $A^0$
and the ratio of the vacuum expectation values (vevs)
of the two Higgs doublets, $\tan\beta = v_2/v_1$.
(For a review of the MSSM Higgs sector, see \cite{HHG}.)
In terms of these parameters, the mass of the charged Higgs boson $H^\pm$
at tree level is
$M^2_{H^\pm} =  M_A^2 + M_W^2$\,,
and the masses of the CP-even neutral Higgs bosons $h^0$ and $H^0$ are obtained
by diagonalizing the tree-level mass-squared matrix,
\begin{Eqnarray}
    \mathcal{M}^2 &=& \left( \begin{array}{cc}
        M_A^2 \sin^2\beta + M_Z^2 \cos^2\beta
            & -(M_A^2 + M_Z^2) \sin\beta \cos\beta \\
        -(M_A^2 + M_Z^2) \sin\beta \cos\beta
            & M_A^2 \cos^2\beta + M_Z^2 \sin^2\beta
        \end{array} \right)\,.
\label{eq:Htreematrix}
\end{Eqnarray}

\vspace{-0.1in}\noindent
The eigenvalues of this matrix are,
\begin{equation}
    M^2_{H^0, h^0} = \frac{1}{2}
    \left[ M_A^2 + M_Z^2 \pm \sqrt{\left(M_A^2 + M_Z^2\right)^2 -
           4 M_A^2 M_Z^2 \cos^2 2 \beta }\, \right]\,,
\end{equation}
with $M_{h^0} < M_{H^0}$.  At tree-level, $M_{h^0} \leq M_Z |\cos 2\beta|$;
this bound is saturated at large $M_A$.  We choose a convention where
the vevs are positive so that $0<\beta<\pi/2$.
The mixing angle that diagonalizes $\mathcal{M}^2$ is given at tree-level by
\begin{equation}
 \tan 2 \alpha = \tan 2 \beta\, {{M_A^2 + M^2_Z}\over{M_A^2 -M^2_Z}}\,.
\end{equation}
In the conventions employed here, $-\pi/2<\alpha<0$ (see
ref.~\cite{ghii} for further details).
From the above results it is easy to obtain:
\begin{equation}
\cos^2(\beta-\alpha)={M_{h^0}^2(M_Z^2-M_{h^0}^2)\over
M_A^2(M_{H^0}^2-M_{h^0}^2)}\,.
\label{cbmasq}
\end{equation}
In the limit of $M_A\gg M_Z$, the expressions for the
Higgs masses and mixing angle simplify and one finds
\begin{Eqnarray}
M_{h^0}^2 &\simeq &\ M_Z^2\cos^2 2\beta\,,\nonumber \\[3pt]
M_{H^0}^2 &\simeq &\ M_A^2+M_Z^2\sin^2 2\beta\,,\nonumber \\[3pt]
\cos^2(\beta-\alpha)&\simeq &\ {M_Z^4\sin^2 4\beta\over 4M_A^4}\,.
\label{largema}
\end{Eqnarray}

\vspace{-0.1in}\noindent
Two consequences are immediately apparent.
First, $M_A\simeq M_{H^0}\simeq M_{H^\pm}$,
up to corrections of ${\mathcal O}(M_Z^2/M_A)$.  Second,
$\cos(\beta-\alpha)=0$ up to corrections of ${\cal O}(M_Z^2/M_A^2)$.
This limit is known as the decoupling limit because when $M_A$ is
large, one can define an effective low-energy theory below the scale
of $M_A$ in which the effective Higgs sector consists only of one light
CP-even Higgs boson, $h^0$, whose couplings to Standard Model particles
are indistinguishable from those of the SM Higgs boson
\cite{decoupling}.
From eq.\ \ref{largema}, one can easily derive:
\begin{equation}
    \cot\alpha = -\tan\beta - \frac{2M_Z^2}{M_A^2} \tan\beta \cos
2\beta
    + \mathcal{O}\left(\frac{M_Z^4}{M_A^4}\right)\,.
\label{eq:cotalphaexpansion}
\end{equation}

When radiative corrections to the CP-even Higgs mass-squared matrix
are taken into account, the upper bound on $M_{h^0}$
increases substantially to $M_{h^0} \lsim 135$ GeV
(assuming all supersymmetric particles are no heavier than about
1~TeV), and corrections to $\alpha$ become substantial for low $M_A$.
These corrections are well known
\cite{hmasseffective,hmass2loopLL,ez,hmassFD1loop,pbmz,hmassFD2loop,chhhww}
and the leading contributions have been computed up to two-loop order.
In this paper we consider only the contributions to the $h^0 b \bar b$
coupling of order $\alpha_s$ at one loop.  Because
the $\mathcal{O}(\alpha_s)$
contributions to the CP-even Higgs mass-squared matrix only first arise
at the two-loop level, the radiative corrections to this matrix are
irrelevant
to our present work.  (In contrast, they do contribute to the one-loop
SUSY-electroweak corrections to the $h^0 b \bar b$ coupling.)

From direct searches at LEP the MSSM $h^0$ and $A^0$ masses
are constrained to be $M_{h^0} > 88.3$ GeV and $M_A > 88.4$ GeV
\cite{LEPhiggs}.
For a range of values of $\tan\beta$ close to one, the theoretical
upper bound on $M_{h^0}$ is lower than the experimental lower bound,
so the corresponding region of $\tan\beta$ can be ruled out.  Because of
the radiative corrections, the variation of
the upper bound depends primarily on the precise value of the top
quark mass and the mixing in the stop sector.
For the conservative choice of
$m_t < 179.4$ GeV and mixing in the stop sector that maximizes the
upper bound on $M_{h^0}$, values of $\tan\beta$ between 0.8 and 1.5
are excluded \cite{LEPhiggs}.

We now discuss the parameters of the sbottom sector.
The tree-level sbottom squared-mass matrix is:
\begin{equation}
    \mathcal{M}^2_{\tilde b} =
    \left(\begin{array}{cc}
    M_L^2 & m_b X_b \\
    m_b X_b & M_R^2 \end{array} \right)\,,
\label{eq:sbottommatrix}
\end{equation}
where we use the notation,
\begin{Eqnarray}
    X_b &=& A_b - \mu \tan\beta\,, \nn \\
    M_L^2 &=& M_{\tilde Q}^2 + m_b^2
    + M_Z^2 (I_3^b - Q_b s^2_W) \cos 2\beta\,,  \nn \\
    M_R^2 &=& M_{\tilde D}^2 + m_b^2 + M_Z^2 Q_b s^2_W \cos 2\beta\,.
    \label{eq:sbottomparams}
\end{Eqnarray}

\vspace{-0.1in}\noindent
Here $I_3^b = -1/2$ and $Q_b = -1/3$ are the isospin and electric
charge of the $b$-quark, respectively and $s_W\equiv \sin\theta_W$.  The
parameters
$M_{\tilde Q}$ and $M_{\tilde D}$ are the soft-SUSY-breaking masses
for the third-generation SU(2) squark doublet $\tilde Q=(\widetilde
t_L, \widetilde b_L)$
and the singlet $\tilde D=\widetilde b_R^\ast$, respectively.
$A_b$ is a soft-SUSY-breaking trilinear
coupling and $\mu$ is the bilinear coupling of the two Higgs doublet
superfields. The sbottom mass eigenstates are
\begin{equation}
    \tilde b_1 = \cos \theta_{\tilde b} \, \tilde b_L
    + \sin \theta_{\tilde b} \, \tilde b_R\,;
    \qquad
    \tilde b_2 = -\sin \theta_{\tilde b} \, \tilde b_L
    + \cos \theta_{\tilde b} \, \tilde b_R\,,
\end{equation}
where $-\pi/4 \leq \theta_{\tilde b} \leq \pi/4$ is defined so that
$\tilde b_1$ ($\tilde b_2$) is predominantly $\tilde b_L$ ($\tilde b_R$).
The sbottom mass eigenvalues are then given by
\begin{equation}
    M^2_{\tilde b_{1,2}} = \frac{1}{2}\left[ M_L^2 + M_R^2
    \pm \sigma_{LR}
    \sqrt{ (M_L^2 - M_R^2)^2 + 4 m_b^2 X_b^2 } \right]\,,
\end{equation}
where\footnote{If $M_L=M_R$, then $\sigma\ls{LR}$ is not well-defined.
In the present context, a useful convention is to set
$\sigma\ls{LR}=\sigma\ls{X}$ [where $\sigma\ls{X}\equiv {\rm
sgn}(X_b)$] if $M_L=M_R$.  Nevertheless, one can check that our final
expressions for the radiative corrections in Section \ref{sec:analytic} 
are independent of this choice.}
\begin{equation}
\sigma\ls{LR}\equiv {\rm sgn}(M_L^2 - M_R^2)\,,
\end{equation}
and the mixing angle $\theta_{\tilde b}$ is given by
\begin{Eqnarray} \label{thetab}
    \cos 2 \theta_{\tilde b} &=& \frac{M_L^2 - M_R^2}
    {M^2_{\tilde b_1}-M^2_{\tilde b_2}}, \nonumber \\
    \sin 2 \theta_{\tilde b} &=&
    \frac{2 m_b X_b}{M^2_{\tilde b_1}-M^2_{\tilde b_2}}\,.
\end{Eqnarray}

\vspace{-0.1in}\noindent
Note that in our conventions, $M_{\tilde b_1}>M_{\tilde b_2}$ if
$\sigma\ls{LR}>0$, whereas the order of the sbottom masses switches if
$\sigma\ls{LR}<0$.

From direct searches at the Tevatron \cite{Tevatronsquark}, the sbottoms
must be heavier than about 140 GeV, assuming that the mass of the lightest
neutralino $\tilde \chi^0_1$ is less than half the mass of the lighter
sbottom.
For heavier neutralinos, the Tevatron searches lose efficiency.
In this region the
direct searches at LEP \cite{LEPsquark} place a lower bound on the sbottom
masses of about 85 GeV.  
The LEP bounds are valid only for $\tilde b - \tilde \chi^0_1$
mass splittings larger than about 5 GeV, so that the decay mode
$\tilde b \to b \tilde \chi^0_1$ is kinematically accessible.

The limits on the gluino mass $M_{\tilde g}$ are more model-dependent.  If one
assumes relations between the gaugino masses such that they unify at
the GUT scale, then $M_{\tilde g}$ is constrained from direct
searches at the Tevatron to be greater than 173 GeV, independent of the
squark masses \cite{Tevatronsquarkgluino}.

\section{SQCD corrections to $h^0 \to b \bar{b}$}
\label{sec:exactcorr}
The $h^0 b \bar{b}$ coupling is given at one-loop level to order
$\alpha_s$ by
\begin{equation}
    \bar{g}_{hbb} = g_{hbb} +
    \delta g_{hbb}^{QCD}+\delta g_{hbb}^{SQCD}\equiv
        g_{hbb}\left(1+\Delta_{QCD}+\Delta_{SQCD}\right)\,,
\end{equation}
where $\bar{g}_{hbb}$ is the one-loop coupling, $g_{hbb}$ is
the tree-level coupling, $\delta g_{hbb}^{QCD}$ is the radiative
correction from pure QCD \cite{qcd}, and $\delta g_{hbb}^{SQCD}$ is the
one-loop SQCD contribution.

The tree-level $h^0 b \bar b$ coupling is given by
\begin{equation} \label{hbbtree}
    g_{hbb} = \frac{gm_b \sin\alpha}{2M_W \cos\beta}\,.
\end{equation}
Note that in the limit of large $M_A$, $\sin\alpha \to - \cos\beta$
and $g_{hbb}$ tends to the SM coupling,
$g_{hbb}^{SM}=-{gm_b}/(2M_W)$.
The one-loop corrections to the $h^0 b \bar b$ coupling modify the
$h^0 \to b \bar b$ decay width as follows, keeping only correction terms
of ${\mathcal{O}}(\alpha_s)$:
\begin{equation}
    \bar{\Gamma}(h^0 \to b \bar b) = \Gamma(h^0 \to b \bar b)
    (1 + 2 \Delta_{QCD} + 2 \Delta_{SQCD})\,,
\end{equation}
where $\bar{\Gamma}$ is the one-loop partial width and $\Gamma$
is the tree-level partial width.

The SQCD contribution to the $h^0 b \bar{b}$ coupling
comes from diagrams involving the
exchange of virtual gluinos ($\tilde{g}$) and sbottoms ($\tilde{b_i}$),
as shown in fig.\ \ref{fig:fd}.
\begin{figure}
\begin{center}
\resizebox{10cm}{!}
{\includegraphics*[122,443][424,687]{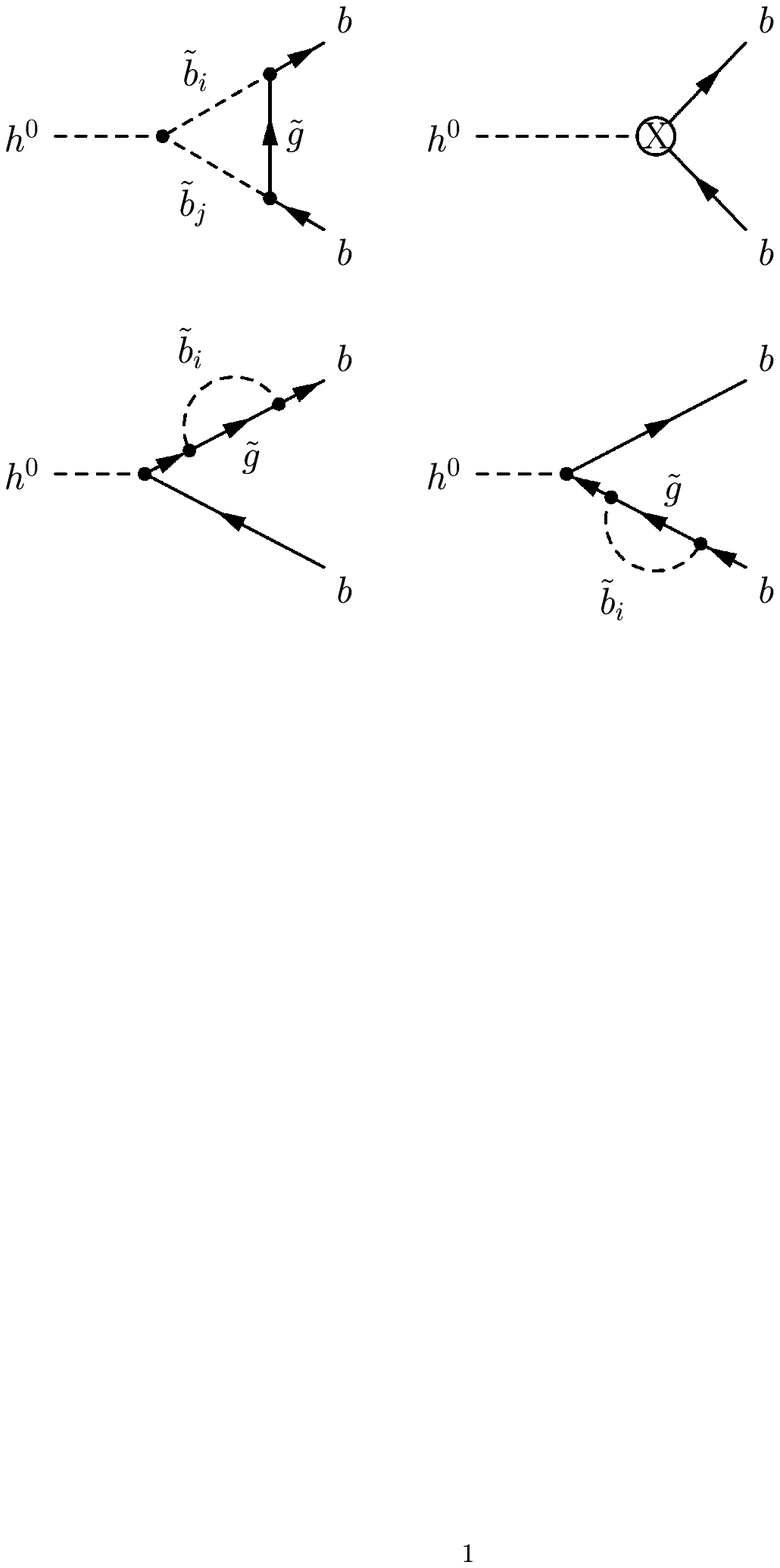}}
\caption{Feynman diagrams for the SQCD corrections to the $h^0 b \bar b$
coupling.  The vertex marked with the $X$ refers to the
one-loop $h^0b\bar b$ counterterm.}
\label{fig:fd}
\end{center}
\end{figure}
We have
\begin{equation} \label{threeterms}
    \delta g_{hbb}^{SQCD} =
    \left(\delta g_{hbb}\right)_{3}^{SQCD} +
    \left(\delta g_{hbb}\right)_{2}^{SQCD} +
    \left(\delta g_{hbb}\right)_{\rm X}^{SQCD}\,,
\end{equation}
consisting of contributions from the vertex correction, the $b$-quark
wave function
renormalization, and the counterterm from the renormalization of
the $b$-quark Yukawa coupling, respectively.
To compute the one-loop Yukawa counterterm contribution,
we note that the Higgs wave function, the vevs
(and hence $\tan\beta$) and the parameters $g$, $M_W$
and $\alpha$ receive no ${\mathcal O}(\alpha_s)$ corrections at
one-loop.  Thus, to leading order in $\alpha_s$,
$\left(\delta g_{hbb}\right)_{\rm X}^{SQCD}$ can be easily obtained
from eq.\ \ref{hbbtree} and
depends only on the $b$-quark mass counterterm as follows:
\begin{equation} \label{mbcounter}
\left(\delta g_{hbb}\right)_{\rm X}^{SQCD} = g_{hbb} \,
{(\delta m_b)^{SQCD} \over m_b}\,.
\end{equation}
In eq.\ \ref{mbcounter}, $(\delta m_b)^{SQCD}$ is the SQCD
contribution to the $b$-quark mass counterterm, which
is fixed by defining $m_b$ to be
the pole of the one-loop ${\mathcal O}(\alpha_s)$ $b$-quark propagator.
This is the on-shell renormalization scheme.

We have computed the various contributions to $\delta g_{hbb}^{SQCD}$
[see eq.\ \ref{threeterms}].  The contribution of the one-loop vertex is
given by:
\begin{Eqnarray}
 &&  \!\!\! \frac{\left(\delta g_{hbb}\right)^{SQCD}_{3}}
    {g_{hbb}}
    = \frac{\alpha_s}{3 \pi}
    \Biggl\{ \left[ \frac{2 M_Z^2}{m_b}
    \frac{\cos\beta \sin(\alpha + \beta)}{\sin\alpha}
    (I_3^b c^2_b - Q_b s^2_W c_{2b})
    + 2 m_b + Y_b s_{2b} \right]  \nn \\[4pt]
    & &
    \qquad\qquad\qquad\qquad \times \left[ m_b C_{11}
    + M_{\tilde g} s_{2b} C_0 \right]
    (m_b^2,M_{h^0}^2,m_b^2;M_{\tilde g}^2,M_{\tilde b_1}^2,
    M_{\tilde b_1}^2)
        \nonumber \\[5pt]
 && \qquad\qquad   + \left[ \frac{2 M_Z^2}{m_b}
    \frac{\cos\beta \sin(\alpha + \beta)}{\sin\alpha}
    (I_3^b s^2_b + Q_b s^2_W c_{2b})
    + 2 m_b - Y_b s_{2b} \right] \nonumber \\[4pt]
    & &
    \qquad\qquad\qquad\qquad\times \left[ m_b C_{11}
    - M_{\tilde g} s_{2b} C_0 \right]
    (m_b^2,M_{h^0}^2,m_b^2;M_{\tilde g}^2,M_{\tilde b_2}^2,
    M_{\tilde b_2}^2)
        \nonumber \\[5pt]
  && \qquad\qquad  + \left[ -\frac{M_Z^2}{m_b}
    \frac{\cos\beta \sin(\alpha + \beta)}{\sin\alpha}
    (I_3^b - 2 Q_b s^2_W) s_{2b}
    + Y_b c_{2b} \right]  \nonumber \\[4pt]
    & &
    \qquad\qquad\qquad\qquad\times \left[ 2 M_{\tilde g} c_{2b} C_0
    (m_b^2,M_{h^0}^2,m_b^2;M_{\tilde g}^2,M_{\tilde b_1}^2,
    M_{\tilde b_2}^2)
    \right] \Biggr\}\,,
    \label{eqn:fulldgvertex}
\end{Eqnarray}

\vspace{-0.1in}\noindent
where $c_b\equiv\cos\theta_{\tilde b}$, $c_{2b}\equiv
\cos 2\theta_{\tilde b}$, $s_b\equiv\sin\theta_{\tilde b}$,
{\it etc.}, and
$Y_b$ arises in the Higgs coupling to sbottoms:
\begin{equation}
    Y_b \equiv A_b + \mu \cot\alpha\,.
\end{equation}

The contribution from the $b$-quark self-energy and the $h^0 b\bar
b$ vertex counterterm is given by
\begin{Eqnarray}
  &&  \frac{\left(\delta g_{hbb}\right)^{SQCD}_{2}
    + \left(\delta g_{hbb}\right)^{SQCD}_{\rm X}}{g_{hbb}}
    = \nn \\[5pt]
&&\qquad\quad - \frac{\alpha_s}{3\pi}
    \Biggl\{ \frac{M_{\tilde g}}{m_b} s_{2b}
    \left[B_0 (m_b^2;M_{\tilde g}^2,M_{\tilde b_1}^2)
    - B_0 (m_b^2;M_{\tilde g}^2,M_{\tilde b_2}^2) \right]
     \nn \\[5pt]
    & & \qquad\quad
    - 2 m_b^2 \left[B_1^{\prime}(m_b^2;M_{\tilde g}^2,M_{\tilde b_1}^2)
    + B_1^{\prime}(m_b^2;M_{\tilde g}^2,M_{\tilde b_2}^2) \right]
    \nonumber \\[5pt]
    & & \qquad\quad
     - 2 m_b M_{\tilde g} s_{2b}
    \left[B_0^{\prime}(m_b^2;M_{\tilde g}^2,M_{\tilde b_1}^2)
    - B_0^{\prime}(m_b^2;M_{\tilde g}^2,M_{\tilde b_2}^2) \right]
    \Biggr\}\,.
    \label{eqn:fulldgwfct}
\end{Eqnarray}

\vspace{-0.1in}\noindent

Our notation for the loop integrals $B_0$, $B_0^{\prime}$, $B_1^{\prime}$,
$C_0$ and $C_{11}$ is defined in the Appendix.
We have checked that our results are
in agreement with the calculations of ref.~\cite{Dabelstein}.

\section{Analytic and numerical results}
\label{sec:analytic}
We now analyze the decoupling behavior
of the SQCD corrections to the $h^0 b \bar b$ coupling.
We derive approximate analytic expressions for the SQCD corrections
in the limit of large SUSY mass parameters and explore the nature
of the decoupling limit.

We define our expansion parameters as follows.  Since we are interested
in the limit of large SUSY mass parameters,
we consider all the soft-SUSY-breaking
mass parameters and the $\mu$ parameter to be of the same order
(collectively denoted by $M_{SUSY}$) and much heavier than the
electroweak scale.  That is,
\begin{equation}
    M_{SUSY} \sim M_L \sim M_R \sim M_{\tilde g} \sim \mu \sim A_b
    \gg M_{EW},
\end{equation}
where $M_L$ and $M_R$ are defined in eq.\ \ref{eq:sbottomparams}.
We give expansions of the SQCD corrections to
the $h^0 b \bar b$ coupling in inverse powers of the SUSY mass parameters,
up to order $M_{EW}^2/M^2_{SUSY}$.  We consider $M_Z$, $M_{h^0}$,
$m_b \tan\beta$, and $m_b \cot\alpha$ to all be of order $M_{EW}$.
We neglect small contributions of order $m_b^2/M^2_{SUSY}$ and
$m_b M_{EW} / M^2_{SUSY}$ that are
not enhanced by $\tan\beta$ or $\cot\alpha$.
The expansions of the loop integrals are given in the Appendix.
There are two possible extreme
configurations of the sbottom mass-squared matrix that we must
separately consider:
maximal and near-zero mixing.

Maximal mixing ($\theta_{\tilde b} \simeq \pm \pi/4$)
between $\tilde b_L$ and $\tilde b_R$
arises when the splitting between
the diagonal elements of the mass-squared matrix is small compared
to the off-diagonal elements: $|M_L^2 - M_R^2| \ll m_b |X_b|$.
Because of the $\tan\beta$ enhancement in $X_b$, $m_b X_b$ is of order
$M_{EW} M_{SUSY}$.  In this case we consider $|M_L^2 - M_R^2|$ to be
of order $M_{EW}^2$, so that
the mass splitting between the two sbottoms is small compared to
their masses and we must take care to treat it properly in the expansions.
We consider this case in Section \ref{sec:45deg}.

Near-zero mixing between $\tilde b_L$ and $\tilde b_R$ arises when the
splitting between the diagonal elements of the mass-squared matrix
is large compared to the off-diagonal elements,
$|M_L^2 - M_R^2| \gg m_b |X_b|$.  This is the case usually considered
in the literature, because $M_L$ and $M_R$ depend on two different
soft-SUSY-breaking parameters $M_{\tilde Q}$ and $M_{\tilde D}$,
respectively, and the $b$-quark mass in the off-diagonal elements is small.
In this case the mass splitting between the two sbottoms will be of the same
order as their masses ({\it i.e.}, $|M_L^2 - M_R^2|$ is of order $M_{SUSY}^2$)
and this has to be treated properly in the expansions.
We consider this case in Section \ref{sec:0deg}.

\subsection{Maximal $\tilde b_L - \tilde b_R$ mixing}
\label{sec:45deg}
Maximal mixing in the sbottom sector arises when
$|M_L^2 - M_R^2| \ll m_b |X_b|$.
In this limit, we can expand
the sbottom mass-squared eigenvalues in powers of the small parameter
$(M_L^2 - M_R^2)/m_b X_b$ (which is of order $M_{EW}/M_{SUSY}$)
as follows:
\begin{equation}
   M^2_{\tilde b_{1,2}} \simeq M_S^2 \pm \Delta^2\,,
\end{equation}
where we have defined
\begin{Eqnarray}
    M_S^2 &=& \nicefrac{1}{2}(M_L^2 + M_R^2)
    = \nicefrac{1}{2}(M_{\tilde b_1}^2 + M_{\tilde b_2}^2)  \nn \\[5pt]
    \Delta^2 &=& \sigma\ls{LR} m_b |X_b|
    \left[1 + \frac{(M_L^2 - M_R^2)^2}{8 m_b^2 X_b^2}\right]\,.
    \label{eq:MSandDelta}
\end{Eqnarray}

\vspace{-0.1in}\noindent
Here $M_S^2$ is of order $M^2_{SUSY}$ while $\Delta^2$ is
of order $M_{EW} M_{SUSY}$.  Expanding the expressions for the mixing
angle in terms of the same small parameter, we obtain
\begin{Eqnarray}
    \cos 2 \theta_{\tilde b} &\simeq&
    \left| \frac{M_L^2 - M_R^2}{2 m_b X_b} \right|\,, \nn \\
    \sin 2 \theta_{\tilde b} &\simeq&
    \sigma\ls{LR}\sigma\ls{X}
    \left[1 - \frac{(M_L^2 - M_R^2)^2}{8 m_b^2 X_b^2} \right]\,,
\end{Eqnarray}

\vspace{-0.1in}\noindent
where $\sigma\ls{X}\equiv {\rm sgn}(X_b)$.
Expanding eqs.\ \ref{eqn:fulldgvertex} and \ref{eqn:fulldgwfct} to order
$M^2_{EW}/M_{SUSY}^2$, we find
\begin{Eqnarray}
 &&\hspace{-1cm}   \Delta_{SQCD} = \frac{\alpha_s}{3\pi}
    \left\{ \frac{-\mu M_{\tilde g}}{M_S^2}
    \left( \tan\beta + \cot\alpha \right)
    f_1(R)
    - \frac{Y_b M_{\tilde g} M_{h^0}^2}{12 M_S^4} f_4(R)
    \right.
    \nonumber \\[5pt]
   && \hspace{-0.7cm} +
    \frac{\mu^2 m_b^2 \tan^2\beta}{2 M_S^4}
    \left[ \frac{\cot\alpha}{\tan\beta} f_4(R)
    - \frac{M_{\tilde
    g}}{M_S^2}\left(Y_b-2A_b\frac{\cot\alpha}{\tan\beta}\right) f_3(R) \right]
	\nn \\[5pt]
    && \hspace{-0.7cm} + \left. 
	\frac{2}{3} \frac{M_Z^2}{M_S^2}
    \frac{\cos\beta \sin(\alpha + \beta)}{\sin\alpha}
    I_3^b
    \left( f_5(R) + \frac{M_{\tilde g} X_b}{M_S^2} f_2(R) \right)
    + \mathcal{O} \left( \frac{m_b M_{EW}}{M_{SUSY}^2} \right) \right\},
    \label{eq:45degexpansion}
\end{Eqnarray}

\vspace{-0.1in}\noindent
where $R\equiv  M_{\tilde g}/M_S$.  
The functions $f_i(R)$ arise from the expansions of the loop
integrals and are given in the Appendix.
They are normalized so that $f_i(1) = 1$.  Note that terms of order
$(M_L^2-M_R^2)^2/(m_b^2 X_b^2)$ cancel exactly in the leading order of the
large $M_{SUSY}$ expansion [eq.\ \ref{eq:45degexpansion}].

The first term in eq.\ \ref{eq:45degexpansion} is
zeroth order in $M_{SUSY}$.  That is, if the ratios between the SUSY
parameters are fixed and the SUSY mass scale is taken arbitrarily heavy,
this term remains constant.  This non-decoupling behavior has been
pointed out previously in refs.~\cite{cmwpheno,cmwcomplementarity}.  If
the
SUSY mass scale is much larger than $M_A$, then one may define a
low-energy
effective theory by integrating out the SUSY particles.  This low-energy
effective theory contains two Higgs doublets, whose couplings to fermions
are  unrestricted ({\it i.e.}, each Higgs doublet couples to
{\it both} up-type and down-type quarks), characteristic of the
so-called general type-III model
\cite{typeIII}.

The remaining terms are of order $M^2_{EW}/M_{SUSY}^2$.  In contrast to the
first term, they depend on $A_b$ (through $X_b$ and $Y_b$).
However, the contribution proportional
to $A_b$ is not enhanced when $\tan\beta$ (or $\cot\alpha$) is large,
and so is less significant at large $\tan\beta$ than the contribution
proportional to $\mu$.
Neglecting all terms that are not enhanced by large $\tan\beta$ or
$\cot\alpha$, we find that $\Delta_{SQCD}$ is proportional to the product
$\mu M_{\tilde g}$.  Because of this, for large $\tan\beta$
the sign of $\Delta_{SQCD}$ can
be used as a test of the anomaly-mediated SUSY breaking scenario
\cite{AMSB}, which predicts a negative $M_{\tilde g}$ \cite{Graham}.
Of course,
the sign of $\mu$ must be determined from another SUSY process for the
sign of $M_{\tilde g}$ to be extracted.

\subsection{Near-zero $\tilde b_L - \tilde b_R$ mixing}
\label{sec:0deg}
Near-zero mixing in the sbottom sector arises when
$|M_L^2 - M_R^2| \gg m_b |X_b|$.
This corresponds to taking the difference between the physical sbottom
masses to be of the same order as the masses themselves.
In this case we write our results in terms of the physical sbottom
masses and expand in powers of the small parameter
$m_b X_b / (M_{\tilde b_1}^2 - M_{\tilde b_2}^2)$, which we take to be
of order $M_{EW}/M_{SUSY}$.
The mixing angle is then given by eq.\ \ref{thetab}, from which one
easily derives the expansion
\begin{equation}
    \cos 2 \theta_{\tilde b} \simeq
    1 - \frac{2 m_b^2 X_b^2}{(M_{\tilde b_1}^2 - M_{\tilde b_2}^2)^2}.
\end{equation}

Expanding eqs.~\ref{eqn:fulldgvertex} and \ref{eqn:fulldgwfct} to order
$M^2_{EW}/M_{SUSY}^2$, and writing the result in terms of the physical
sbottom masses, we find:
\begin{Eqnarray}
&& \!\!\!  \Delta_{SQCD} = \frac{\alpha_s}{3\pi} \left\{
    \frac{-2 \mu M_{\tilde g}}{M_{\tilde b_1}^2 - M_{\tilde b_2}^2}
    (\tan\beta + \cot\alpha) \, h_1(R_1,R_2)
    + 2 M_{h^0}^2
    \frac{M_{\tilde g} Y_b \, h_2(R_1,R_2)}
    {(M_{\tilde b_1}^2 - M_{\tilde b_2}^2)^2}
    \right. \nonumber \\[5pt]
    &&\quad + 2 M_Z^2 \frac{\cos\beta \sin(\alpha + \beta)}{\sin\alpha}
    \left[(I_3^b - Q_b s^2_W)
    \left( \frac{f_5(R_1)}{3M_{\tilde b_1}^2}
    - \frac{M_{\tilde g} X_b}{M_{\tilde b_1}^2 - M_{\tilde b_2}^2}
    \frac{f_1(R_1)}{M_{\tilde b_1}^2}
    \right. \right. \nonumber \\[5pt]
    && \qquad\qquad\qquad\qquad\qquad\qquad\qquad\qquad\qquad
    \left.
    + \frac{2 M_{\tilde g} X_b \, h_1(R_1,R_2)}
    {(M_{\tilde b_1}^2 - M_{\tilde b_2}^2)^2}
    \right)
    \nonumber \\[5pt]
    && \qquad\qquad\qquad 
    + Q_b s^2_W \left( \frac{f_5(R_2)}{3 M_{\tilde b_2}^2}
    + \frac{M_{\tilde g} X_b}{M_{\tilde b_1}^2 - M_{\tilde b_2}^2}
    \frac{f_1(R_2)}{M_{\tilde b_2}^2}
    \left.
    - \frac{2 M_{\tilde g} X_b \, h_1(R_1,R_2)}
    {(M_{\tilde b_1}^2 - M_{\tilde b_2}^2)^2}
    \right) \right]
    \nonumber \\[5pt]
    &&\quad -
    \frac{2 \mu^2 M_{\tilde g} m_b^2 \tan^2\beta}
     {(M_{\tilde b_1}^2 - M_{\tilde b_2}^2)^2}
     \left(Y_b-2A_b\frac{\cot\alpha}{\tan\beta}\right)
    \left(\frac{f_1(R_1)}{M_{\tilde b_1}^2}+\frac{f_1(R_2)}{M_{\tilde b_2}^2}
    -\frac{4h_1(R_1,R_2)}{M_{\tilde b_1}^2 - M_{\tilde b_2}^2}\right) 
     \nn \\[5pt]
       &&\quad -
    \frac{2 \mu^2 m_b^2 \tan\beta\cot\alpha}
     {M_{\tilde b_1}^2 - M_{\tilde b_2}^2}
      	\left( \frac{f_5(R_1)}{3 M_{\tilde b_1}^2}
    	- \frac{f_5(R_2)}{3 M_{\tilde b_2}^2} \right)
       + \left. \mathcal{O} \left( \frac{m_b M_{EW}}{M_{SUSY}^2} \right)
    \right\}\,,
\label{eq:0degexpansion}
\end{Eqnarray}

\vspace{-0.1in}\noindent
where $R_i \equiv M_{\tilde g}/M_{\tilde b_i}$ ($i=1,2$).
The functions $h_i(R_1,R_2)$ and $f_{1,5}(R_i)$ arise
from the expansions of the loop integrals and are given in the Appendix.

As in the case of maximal sbottom mixing, the first term in eq.\
\ref{eq:0degexpansion} is zeroth order in $M_{SUSY}$.
The remaining terms are of order $M^2_{EW}/M^2_{SUSY}$.  As in the previous
section, if we neglect all terms that are not enhanced by large $\tan\beta$
or $\cot\alpha$, we find that the dependence on $A_b$ drops out and
$\Delta_{SQCD}$ is again proportional to the product $\mu M_{\tilde g}$.

\subsection{The approach to the decoupling limit}
If we take all SUSY mass parameters large at fixed $M_A$
in eqs.\ \ref{eq:45degexpansion} and \ref{eq:0degexpansion}, 
then $\Delta_{SQCD}$ tends to a nonzero constant; {\it i.e.},
the SQCD corrections do {\it not} decouple.
However, we are especially interested in the case where both
$M_{SUSY}$ and $M_A$ are large.  In 
eqs.\ \ref{eq:45degexpansion} and \ref{eq:0degexpansion}, the
terms of zeroth order in $M_{SUSY}$ are proportional to
$\tan\beta + \cot\alpha$.  From eq.\ \ref{eq:cotalphaexpansion},
\begin{equation} \label{tbpca}
    \tan\beta + \cot\alpha
    = - \frac{2M_Z^2}{M_A^2} \tan\beta \cos 2\beta
    + \mathcal{O}\left(\frac{M_{EW}^4}{M_A^4}\right)\,.
\end{equation}
Thus, the first term in eq.\ \ref{eq:45degexpansion} and in
eq.\ \ref{eq:0degexpansion} is of order $M_{EW}^2\tan\beta/M_A^2$, and
therefore decouples
in the limit of large $M_A$.  However, the approach to decoupling is
delayed in the large $\tan\beta$ regime.\footnote{The
enhancement of the radiatively-corrected $h^0 b\bar b$ coupling 
at large $\tan\beta$ has
also been emphasized in refs.~\cite{cjs,cmwpheno,polonsky,kolda,wells2}.}
Specifically, for values of
$M_A^2\sim M_Z^2\tan\beta$, we see that $\tan\beta+\cot\alpha\sim
{\mathcal O}(1)$. For example, if $\tan\beta\sim 50$, then even for
values of $M_A\sim 1$~TeV, decoupling has not yet set in.

Other terms in eqs.\ \ref{eq:45degexpansion} and
\ref{eq:0degexpansion} also exhibit delayed decoupling.
In particular, eq.\ \ref{tbpca} implies that
\begin{equation}
Y_b = X_b +\mathcal{O}\left(\frac{M_{SUSY}M_{EW}^2\tan\beta}
{M_A^2}\right)\,,
\end{equation}
so that $Y_b$ is also enhanced at large
$\tan\beta$. 
Hence, all terms in eqs.\ \ref{eq:45degexpansion} and
\ref{eq:0degexpansion} that are proportional to either $X_b$ or $Y_b$
are of order $M^2_{EW} \tan\beta / M^2_{SUSY}$.  Again,
if $\tan\beta \sim 50$ and $M_{SUSY} \sim 1$~TeV, decoupling has not
yet set in.

The remaining terms in eqs.\ \ref{eq:45degexpansion} and
\ref{eq:0degexpansion} exhibit the expected decoupling in the usual
sense (with no delay).  In particular,
we may set $\alpha=\beta-\pi/2$ in the decoupling
limit to obtain
\begin{equation}
{\cos\beta\sin(\alpha+\beta)\over\sin{\alpha}}=\cos 2\beta+{\mathcal
O}\left({M_{EW}^2\over M_A^2}\right)\,,
\end{equation}
which exhibits no $\tan\beta$ enhancement.  All remaining
factors of $\tan\beta$
are multiplied by the appropriate power of $m_b$, and since
$m_b\tan\beta\sim M_{EW}$, no delayed decoupling results from these
terms.


We have thus shown analytically that the one-loop SQCD corrections to
the $h^0 b \bar b$ coupling decouple in the limit of large
$M_{SUSY}$ and large $M_A$.  The decoupling takes the generic form:
\begin{equation} \label{genform}
    \Delta_{SQCD} \sim C_1\frac{M_{EW}^2}{M_A^2} + C_2
\frac{M_{EW}^2}{M_{SUSY}^2}\,.
\end{equation}
In general $C_1$ approaches a non-zero constant as
$M_{SUSY}\to\infty$, while $C_2$ approaches a (different) non-zero
constant as $M_A\to\infty$.  Thus, the decoupling limit requires both
$M_A$ and $M_{SUSY}$ to become simultaneously large (as compared to
$M_{EW}$). However, we will
demonstrate that in some cases the SQCD radiative corrections vanish in
the limit where some SUSY particle masses are large, independent of the
value of $M_A$.

This decoupling is shown numerically~\footnote{In our numerical analysis
we have taken the $b$-quark pole-mass to be $4.75$~GeV and $\alpha_s = 0.119$.
Because of the experimental constraints on the sbottom masses, 
we consider only regions of parameter space in which both 
sbottoms are heavier than 100 GeV.}
in figs.\ \ref{fig:2cd50plot} and \ref{fig:MSMA}.
\begin{figure}
\begin{center}
\resizebox{10cm}{!}{\rotatebox{270}{\includegraphics{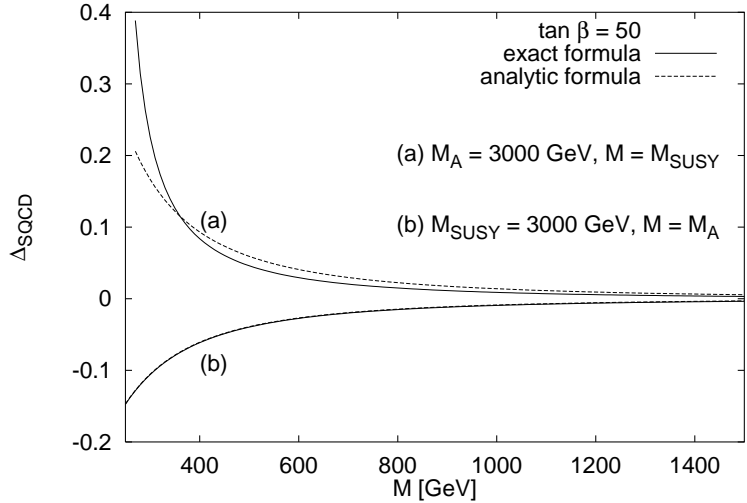}}}
\caption{$\Delta_{\rm SQCD}$ as a function of particle mass for
$\tan\beta = 50$ and $M_{SUSY} = M_L=M_R=M_S = M_{\tilde g} = \mu =
A_b$. The curves (a) are plotted {\it vs.}~$M_{SUSY}$,
with $M_A=3000$~GeV; whereas the curves (b) are plotted 
{\it vs.}~$M_A$, with $M_{SUSY}=3000$~GeV.  
Solid lines are based on the exact one-loop formula and dashed lines are
based on the analytic approximation of eq.\ \ref{eq:45degexpansion}.}
\label{fig:2cd50plot}
\end{center}
\end{figure}
In fig.\ \ref{fig:2cd50plot}, we plot the
exact one-loop expression for $\Delta_{SQCD}$ (solid lines) and the expansion
of eq.\ \ref{eq:45degexpansion} (dashed lines) for $\tan\beta = 50$
and $M_{SUSY} = M_L=M_R=M_S = M_{\tilde g} = \mu = A_b$.
The lines labeled (a) show $\Delta_{SQCD}$ as a function of
$M_{SUSY}$.  We have
fixed $M_A$ very large, $M_A = 3000$ GeV, in order to eliminate the
contribution to $\Delta_{SQCD}$ that decouples at large $M_A$.
We use the exact tree-level formula for $\cot\alpha$ as a function of $M_A$
and $\tan\beta$.
The lines labeled (b) show $\Delta_{SQCD}$ as a function of
$M_A$, where now we have fixed $M_{SUSY}$ to be very large,
$M_{SUSY} = 3000$ GeV,
in order to examine only the contribution to $\Delta_{SQCD}$ that
does not decouple at large $M_{SUSY}$.
We note that for very large $M_{SUSY}$ and $M_A = 1$ TeV, $\Delta_{SQCD}$
is of order $-1\%$ for $\tan\beta = 50$.  We have plotted our results
for $\mu M_{\tilde g}$ positive.  In the approximation of neglecting terms
not enhanced by large $\tan\beta$ or $\cot\alpha$, changing the sign
of $\mu M_{\tilde g}$ simply flips the sign of $\Delta_{SQCD}$.

\begin{figure}
\begin{center}
\resizebox{15cm}{!}{\rotatebox{270}{\includegraphics{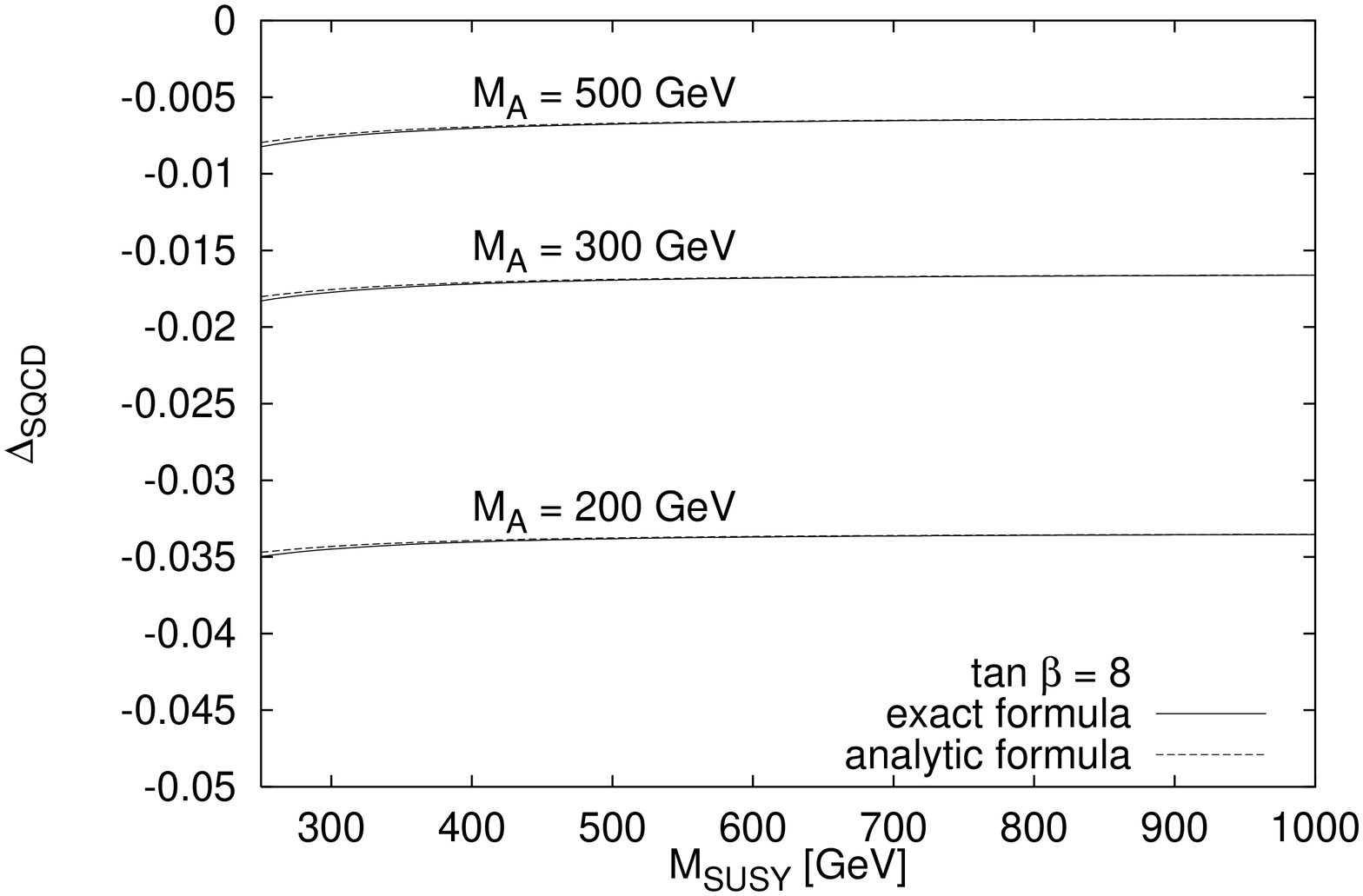}}
\rotatebox{270}{\includegraphics{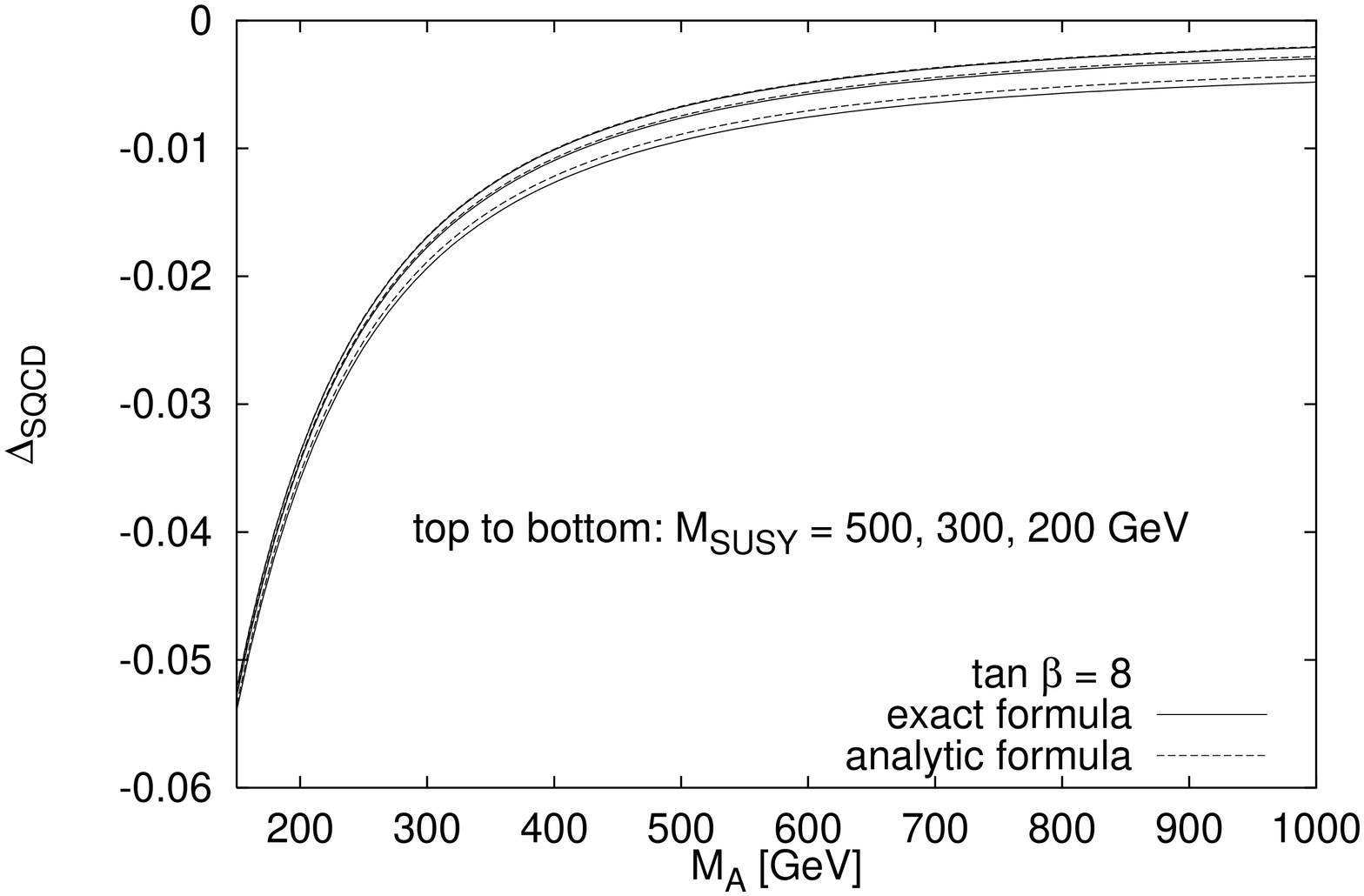}}}
\resizebox{15cm}{!}{\rotatebox{270}{\includegraphics{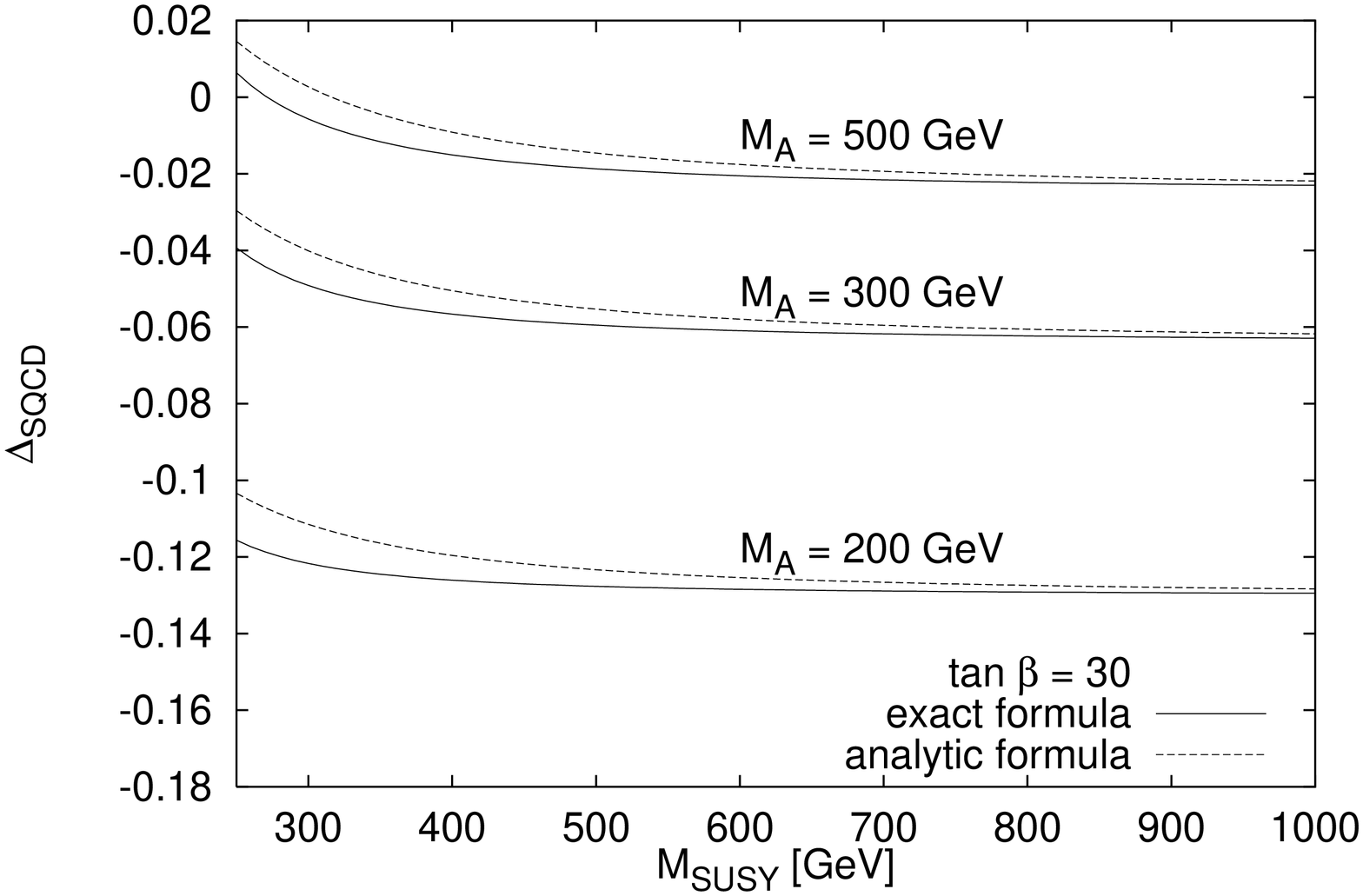}}
\rotatebox{270}{\includegraphics{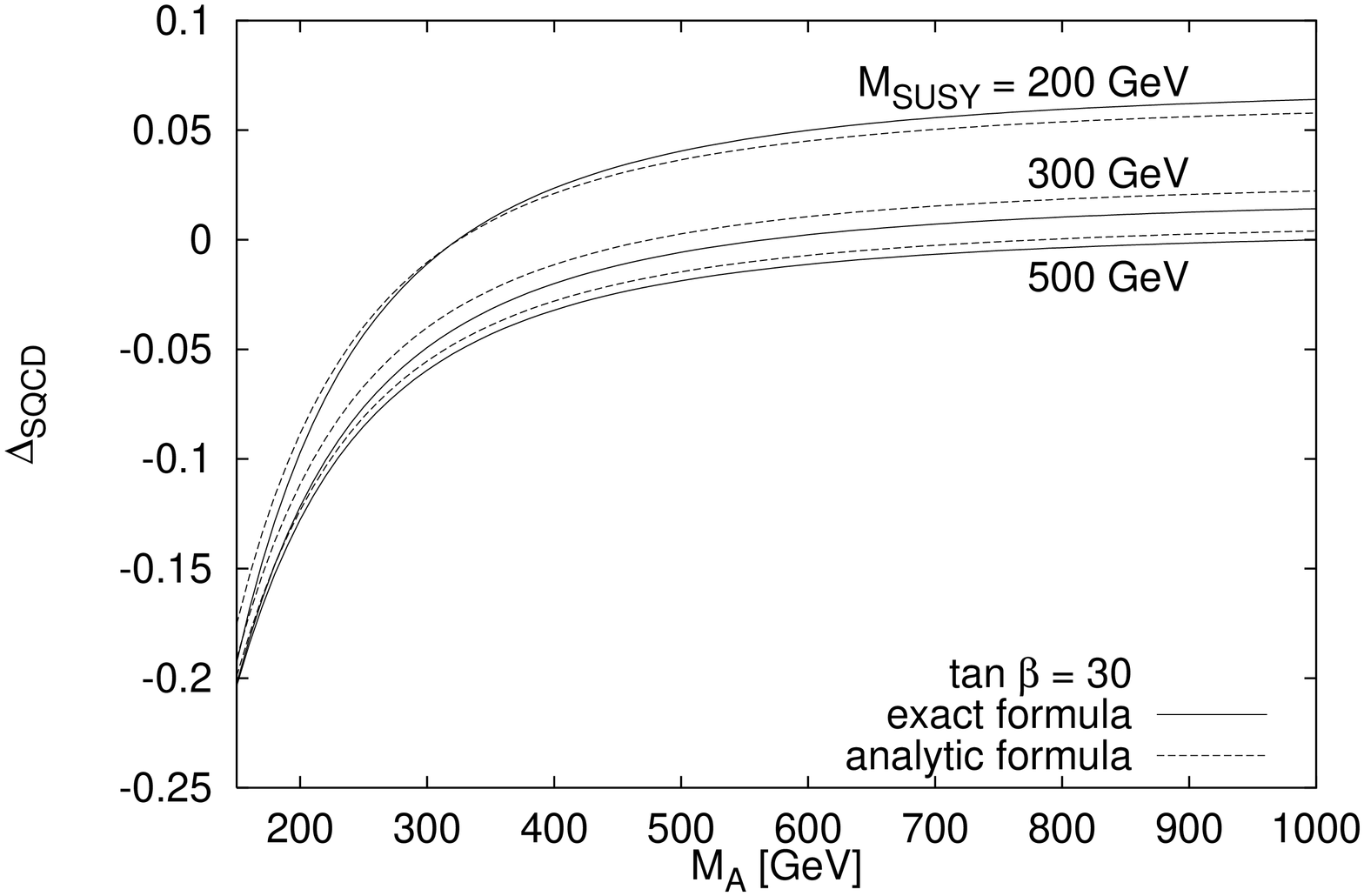}}}
\resizebox{15cm}{!}{\rotatebox{270}{\includegraphics{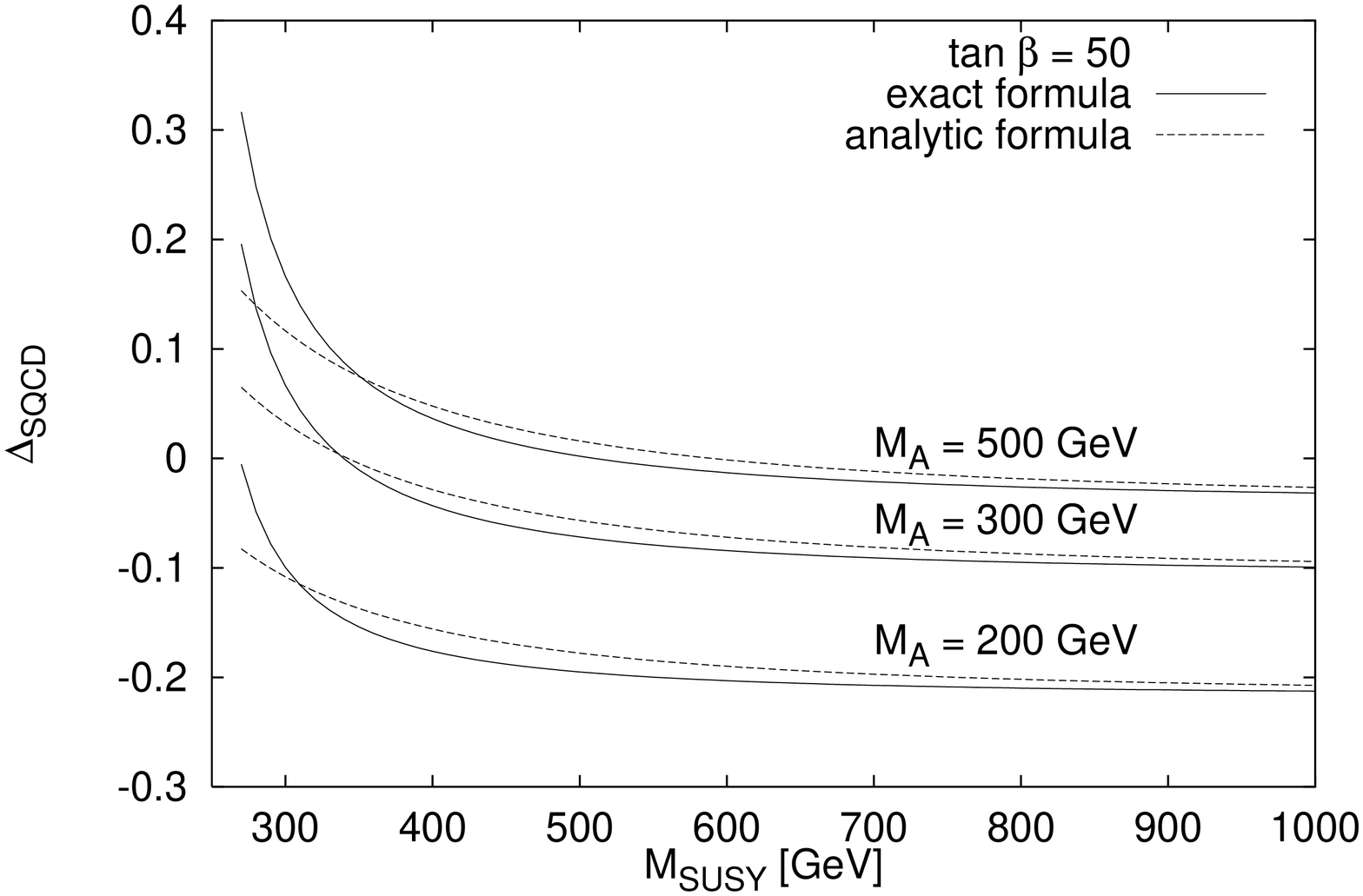}}
\rotatebox{270}{\includegraphics{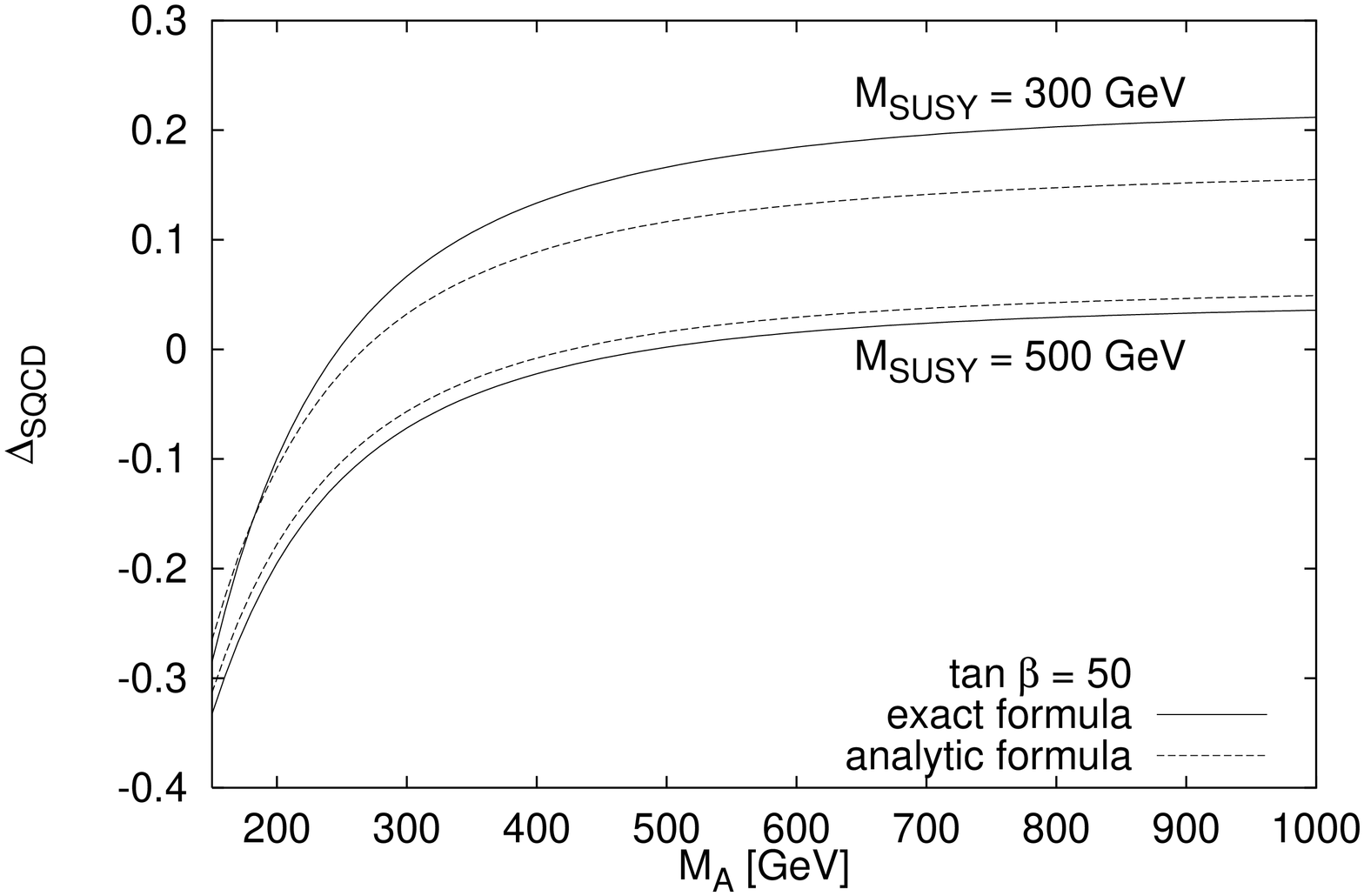}}}
\caption{$\Delta_{SQCD}$ for $M_{SUSY} = M_L=M_R=M_S = M_{\tilde g} =
\mu= A_b$,
with $\tan\beta = 8$ (top panels), 30 (middle panels), and 50 (bottom panels).
The solid lines are based on
the exact one-loop formula and the dashed lines are based on
the analytic approximation of eq.\ \ref{eq:45degexpansion}.
In the left-hand panels we plot $\Delta_{SQCD}$ as a function of $M_{SUSY}$
for $M_A = 200$, 300, and 500 GeV.
In the right-hand panels we plot $\Delta_{SQCD}$ as a function of $M_A$
for $M_{SUSY} = 200$, 300, and 500 GeV.
For $\tan\beta = 50$, the value of
$M_{SUSY}=200$~GeV yields a negative mass-squared
for the lighter sbottom, so this value is not shown in the bottom right
panel.}
\label{fig:MSMA}
\end{center}
\end{figure}
In fig.\ \ref{fig:MSMA} we again plot the exact one-loop expression
for $\Delta_{SQCD}$ (solid lines) and the expansion of eq.\
\ref{eq:45degexpansion} (dashed lines) for all the SUSY mass parameters
equal, $M_{SUSY} = M_L=M_R=M_S = M_{\tilde g} = \mu = A_b$, and three values
of $\tan\beta$.~\footnote{Although we have chosen $M_L=M_R$ for
simplicity, our results are not particularly sensitive to this choice
as long as $|M_L^2-M_R^2|\ll m_b|X_b|$ ({\it c.f.} the remarks below
eq.\ \ref{eq:45degexpansion}).}
Note the change in the vertical scale for the plots
with different values of $\tan\beta$.
We show the dependence
of $\Delta_{SQCD}$ on $M_{SUSY}$ (left-hand panels) and $M_A$ (right-hand
panels).  Clearly, in the limit of large $M_{SUSY}$, $\Delta_{SQCD}$
tends to a non-vanishing constant, and this constant tends to zero
in the large $M_A$ limit. 
Similarly, in the limit of large $M_A$, $\Delta_{SQCD}$ tends to a 
non-vanishing constant, and this constant tends to zero in the large
$M_{SUSY}$ limit.

Notice that from the numerical comparison between the exact and analytic
formulae in fig.\ \ref{fig:MSMA}, we can conclude that our expansion is a
good approximation for large enough SUSY mass parameters.  In particular,
it is reasonably accurate for $M_{SUSY}$ larger than 300 GeV.
Also, it is clear that as $\tan\beta$ grows,
not only does $\Delta_{SQCD}$ increase in magnitude, but the agreement
between the exact and analytic formulae becomes worse at low $M_{SUSY}$.
This is due to the
fact that the splitting between the squared masses of the two sbottoms
in the maximal mixing case is proportional to $m_b \tan\beta$, which we
have taken to be of order $M_{EW}$ in our expansion.  As $\tan\beta$
increases, the mass of the lighter sbottom decreases, and the higher
order terms that we have neglected in our expansion become more important.

All numerical results presented so far correspond to $\mu M_{\tilde
g}>0$.  In the case of $\mu M_{\tilde g}<0$, the qualitative features
of $|\Delta_{SQCD}|$ remain unchanged.  From the analytic formulae
derived in this section, one can see that at large $\tan\beta$ the
dominant effect of changing the sign of $\mu M_{\tilde g}$ is 
to change the overall sign of $\Delta_{SQCD}$.  We can illustrate this point 
in the simple limiting case in which all SUSY mass
parameters and $M_A$ are equal.  Simplifying eq.\ \ref{eq:45degexpansion}
in this limit, we end up with a simple
formula for the case of $\mu M_{\tilde g}>0$: 
\begin{Eqnarray}
    \Delta_{SQCD}
    &=& \frac{\alpha_s}{3\pi}
    \left\{ \frac{M_Z^2}{3M_{SUSY}^2} \cos 2\beta
    \left( 7 \tan\beta - 2 \right)
    + \frac{M_{h^0}^2}{12M_{SUSY}^2} \left( \tan\beta - 1 \right)
    \right. \nn \\
    & & \left. \qquad +  \frac{m_b^2 \tan^2\beta}{2M_{SUSY}^2}
    \left( \tan\beta - 4 \right)
    + \mathcal{O}\left( \frac{m_b M_{EW}}{M_{SUSY}^2} \right)
    \right\}\,,
    \label{eq:allmassesequal}
\end{Eqnarray}

\vspace{-0.1in}\noindent
where $M_{SUSY} = M_S = M_{\tilde g} = \mu = A_b = M_A$.  To
obtain the result for $\mu M_{\tilde g}<0$, one
replaces $\tan\beta$ with $-\tan\beta$ in eq.\ \ref{eq:allmassesequal}.
The formula of eq.\ \ref{eq:allmassesequal} is plotted in
fig.\ \ref{fig:2aplot} for three values of $\tan\beta$ and both signs
of $\mu$ (taking $M_{\tilde g}$ to be positive, by convention).
\begin{figure}
\begin{center}
\resizebox{15cm}{!}{\rotatebox{270}
{\includegraphics{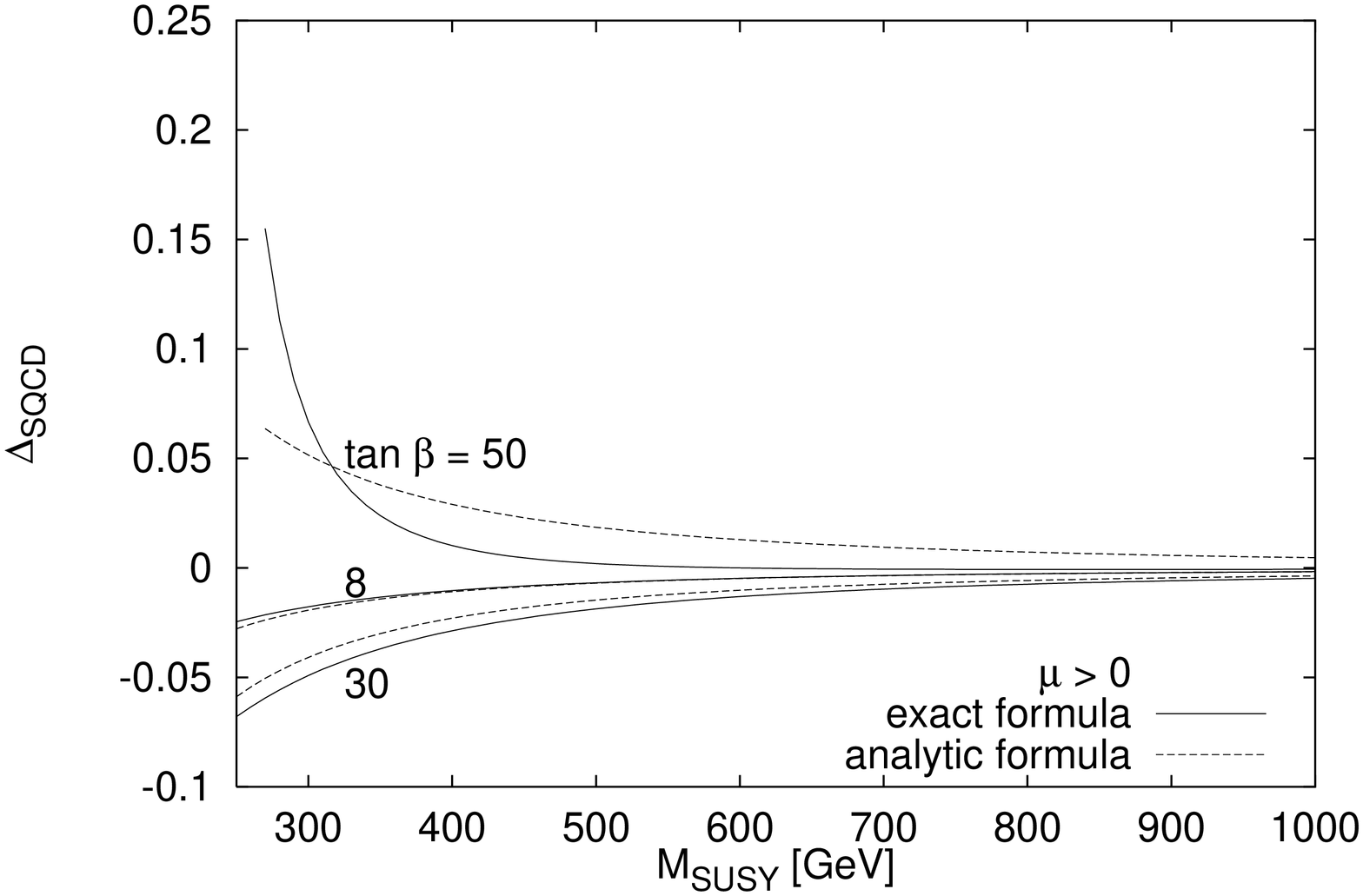}}
\rotatebox{270}{\includegraphics{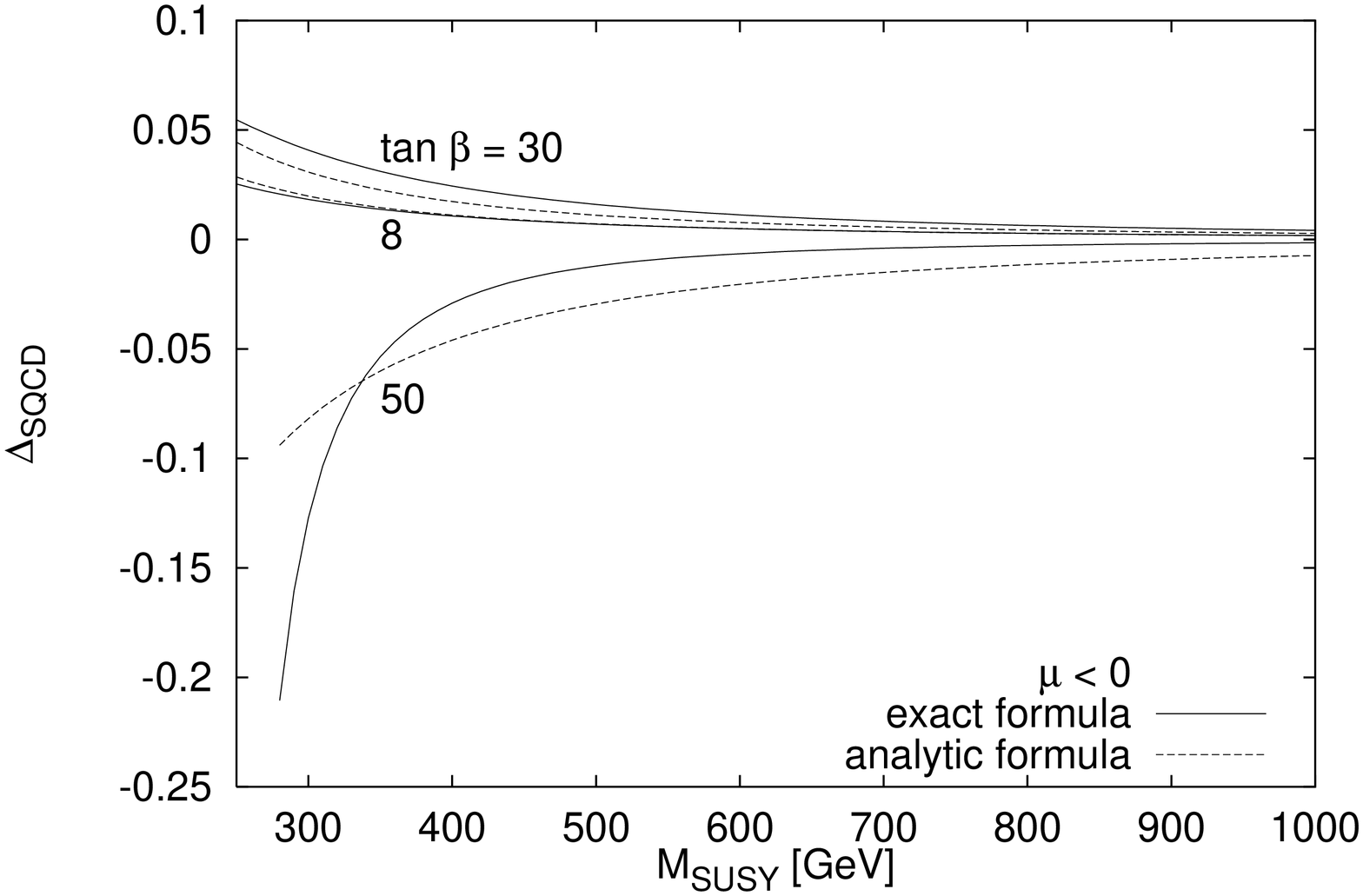}}}
\caption{$\Delta_{\rm SQCD}$ as a function of $M_{SUSY}$ for
$M_{SUSY} = M_L=M_R=M_S = M_{\tilde g} = |\mu| = A_b = M_A$ and
$\tan\beta = 8$, 30, 50.  Both
positive and negative $\mu$ cases are shown.
The solid lines are based on
the exact one-loop formula and the dashed lines are based on
the analytic approximation of eq.\ \ref{eq:allmassesequal}.}
\label{fig:2aplot}
\end{center}
\end{figure}
Clearly, $\Delta_{SQCD}$ decouples like $(M_{EW}^2/M_{SUSY}^2)$,
but this decoupling is delayed at large $\tan\beta$.  For example,
even at $M_{SUSY} = 1$~TeV,
$|\Delta_{SQCD}| \simeq 1$\% for $\tan\beta \sim 30$.
Note that as expected, changing the sign of $\mu$
simply changes the sign of the dominant contribution to $\Delta_{SQCD}$.
In the remainder of our analysis, we will display results only 
for $\mu > 0$.

Next, we consider the decoupling of the SQCD corrections
to the $h^0 b \bar b$ coupling as individual SUSY particles become heavy
compared to the common SUSY mass scale.  We examine three cases:
large $M_S$ with maximal sbottom mixing, large $M_{\tilde g}$ with maximal
sbottom mixing, and one heavy sbottom state with near-zero sbottom
mixing.

We first consider the case of large $M_S$ with maximal sbottom mixing,
with $M_S \gg M_{\tilde g} \sim \mu \sim A_b \gg M_{EW}$.
If $M_S$ is taken large while the rest of the SUSY mass parameters
remain fixed, then we may expand the functions $f_i(R)$ in
eq.\ \ref{eq:45degexpansion} for
$M_S \gg M_{\tilde g}$, or $R \ll 1$.  The result is:
%
\begin{equation}
    \Delta_{SQCD} = \frac{\alpha_s}{3\pi}
    \left\{ \frac{-2 \mu M_{\tilde g}}{M_S^2}(\tan\beta + \cot\alpha)
    +\frac{M_Z^2}{M_S^2}
    \frac{\cos\beta \sin(\alpha + \beta)}{\sin\alpha} I_3^b
    + \mathcal{O}\left(\frac{M^4}{M_S^4}\right) \right\}\,,
    \label{eq:largeMSexp}
\end{equation}
where $M$ is one of the lighter SUSY particle masses.  Note
that in this limit, the SQCD corrections decouple like $M^2/M_S^2$
even for light $M_A$.  Thus
it is only in the case of large $M_{\tilde g}$ and $\mu$, of the
same order as $M_S$, that large $M_A$ is required for decoupling.
\begin{figure}[htbp]
\begin{center}
\resizebox{10cm}{!}{\rotatebox{270}{\includegraphics{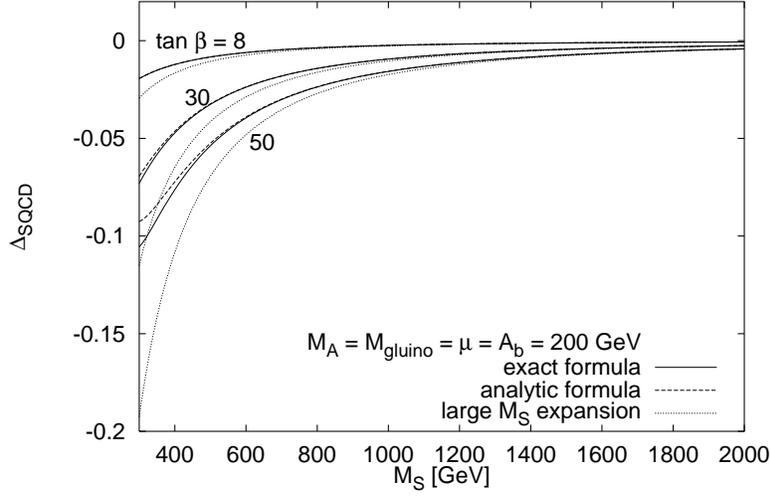}}}
\caption{$\Delta_{\rm SQCD}$ as a function of $M_S$
(assuming $M_L=M_R=M_S$) for
$M_{\tilde g} = \mu = A_b = M_A = 200$ GeV and $\tan\beta = 8, 30, 50$
(top to bottom).
Solid lines are based on
the exact one-loop expression, dashed lines are based on
the analytic expansion of eq.\ \ref{eq:45degexpansion}, and dotted lines
are based on the large $M_S$ expansion of eq.\ \ref{eq:largeMSexp}.}
\label{fig:2eplot}
\end{center}
\end{figure}
\begin{figure}[htbp]
\begin{center}
\resizebox{10cm}{!}{\rotatebox{270}{\includegraphics{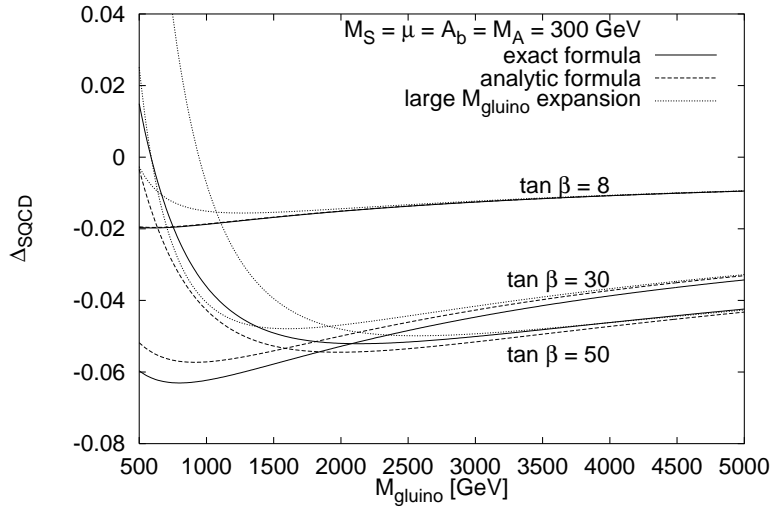}}}
\caption{$\Delta_{\rm SQCD}$ as a function of $M_{\tilde g}$ for
$M_L=M_R=M_S = \mu = A_b = M_A = 300$ GeV and $\tan\beta = 8, 30, 50$.
Solid lines are based on the exact one-loop expression, dashed lines 
based on are
the analytic expansion of eq.\ \ref{eq:45degexpansion}, and
dotted lines are based on the expansion for large $M_{\tilde g}$ of
eq.\ \ref{eq:heavygluinoexp}.}
\label{fig:1plot200}
\end{center}
\end{figure}
In fig.\ \ref{fig:2eplot} we plot the exact one-loop expression for
$\Delta_{SQCD}$ and the expansions of eqs.\ \ref{eq:45degexpansion}
and \ref{eq:largeMSexp} as
a function of $M_S$, for fixed
$M_{\tilde g} = \mu = A_b = M_A = 200$ GeV and three different values
of $\tan\beta$.  This figure shows the decoupling
of $\Delta_{SQCD}$ with $M_S$ as discussed above.

Similarly we examine the case of a very heavy gluino compared to the
rest of the SUSY spectrum.  We still focus on the case of maximal sbottom
mixing.  Expanding the functions
$f_i(R)$ in eq.\ \ref{eq:45degexpansion} for $M_{\tilde g} \gg M_S$,
or $R \gg 1$, we see that in this case
the SQCD corrections decouple with the gluino mass like $M/M_{\tilde
g}$, where again $M$ is one of the other light SUSY masses:
\begin{Eqnarray}    \label{eq:heavygluinoexp} 
  &&\!\!\!\!  \Delta_{SQCD} = \frac{\alpha_s}{3\pi}
    \left\{ \frac{2 \mu}{M_{\tilde g}}(\tan\beta + \cot\alpha)
    \left[1 - \log\left(\frac{M^2_{\tilde g}}{M_S^2} \right)\right]
    \right.
   - \frac{Y_b}{3M_{\tilde g}} \frac{M_{h^0}^2}{M_S^2}
  \\[5pt]
  &&     +\frac{2X_b}{M_{\tilde g}} \frac{M_Z^2}{M_S^2}
    \frac{\cos\beta \sin(\alpha + \beta)}{\sin\alpha} I_3^b 
   \left.
    - \frac{\mu^2 m_b^2 \tan^2\beta}{M_{\tilde g}M_S^4}\left(Y_b
     -2A_b\frac{\cot\alpha}{\tan\beta}\right)
    + \mathcal{O}\left(\frac{M^2}{M_{\tilde g}^2}\right) \right\}. \nn
\end{Eqnarray}

\vspace{-0.1in}\noindent
Note that the decoupling of the SQCD corrections at large
$M_{\tilde g}$
(with all other SUSY mass parameters held fixed) is very slow:
$\Delta_{SQCD}$ falls off only as one power of $M_{\tilde g}$.  This is due
to the factor of $M_{\tilde g}$ in the numerator of eqs.\
\ref{eqn:fulldgvertex} and \ref{eqn:fulldgwfct},
which arises from the gluino propagator.
$\Delta_{SQCD}$ is also
enhanced by the factor $\log(M_{\tilde g}^2/M_S^2)$.  
We again see the phenomenon of delayed decoupling at large $\tan\beta$ due
to the terms in eq.\ \ref{eq:heavygluinoexp} proportional to 
either $X_b$ or $Y_b$.

In fig.\ \ref{fig:1plot200} we plot the exact one-loop expression for
$\Delta_{SQCD}$ and the expansions of eqs.\ \ref{eq:45degexpansion}
and \ref{eq:heavygluinoexp} as
a function of $M_{\tilde g}$, for 
$M_S = \mu = A_b = M_A = 300$~GeV and three different values
of $\tan\beta$.  This figure shows the slow decoupling
of $\Delta_{SQCD}$ with $M_{\tilde g}$.  
For example, for $M_{\tilde g} = 500$ GeV and
$\tan\beta = 30$, $\Delta_{SQCD}\simeq -6\%$ for
$M_S = \mu = A_b = M_A = 300$ GeV.  If the latter masses are reduced
to 200~GeV, one finds $\Delta_{SQCD}\simeq -13\%$, which is a
significant correction.  Fig.\ \ref{fig:1plot200} also illustrates
the validity of the large gluino mass expansion.  This expansion is
particularly poor for large values of $\tan\beta$ out to a very large
gluino mass of about 2000~GeV.

Finally we study the case in which one of the sbottoms becomes
heavy while the other sbottom mass and the rest of the SUSY mass parameters
are fixed.  We choose $M_R \gg M_L \sim M_{\tilde g} \sim \mu \sim A_b
\gg M_{EW}$, so that $M_{\tilde b_2} \gg M_{\tilde b_1}$.
This is necessarily the case of near-zero sbottom mixing.  Expanding
eq.\ \ref{eq:0degexpansion} in inverse powers of $M_{\tilde b_2}$,
we find:
\begin{Eqnarray}
    \Delta_{SQCD} &=& \frac{\alpha_s}{3\pi} \left\{
    \frac{2}{3} \frac{M_Z^2}{M^2_{\tilde b_1}}
    \frac{\cos\beta \sin(\alpha + \beta)}{\sin\alpha}
    (I_3^b - Q_b s^2_W) f_5(R_1)
    \right. \nonumber \\
    &+& \frac{2 \mu M_{\tilde g}}{M_{\tilde b_2}^2}
    (\tan\beta + \cot\alpha) \left[h(R_1) +
    \log \left(\frac{M_{\tilde g}^2}{M_{\tilde b_2}^2}\right)\right]
    \nonumber \\
    &+& \frac{M_Z^2}{M^2_{\tilde b_2}}
    \frac{\cos\beta \sin(\alpha + \beta)}{\sin\alpha}
    \left[ (I_3^b - Q_b s^2_W)
    \frac{2 M_{\tilde g} X_b}{M_{\tilde b_1}^2} f_1(R_1)
    + Q_b s^2_W \right]
    \nn \\
    &+& \left.
    \frac{2}{3} \frac{\mu^2 m_b^2 \tan\beta \cot\alpha}{M_{\tilde b_1}^2
    M_{\tilde b_2}^2} f_5(R_1)
    + \mathcal{O}\left(\frac{M^4}{M_{\tilde b_2}^4}\right) \right\}\,,
    \label{eq:heavyb2}
\end{Eqnarray}

\vspace{-0.1in}\noindent
where again $M$ is one of the other light SUSY masses and the function
$h(R_1)$ is given in the Appendix.  Note that
the first term does {\it not} decouple as $M_{\tilde b_2}$ is taken
large.  This behavior is independent of the value of $M_A$ (and
therefore holds even if $M_A\to\infty$). However, this first term is not
enhanced by large
$\tan\beta$ and is numerically negligible as can be seen in fig.\
\ref{fig:0deg:1aplot}.  The terms that are
enhanced by large $\tan\beta$ decouple like $M^2/M_{\tilde b_2}^2$.
In fig.\ \ref{fig:0deg:1aplot} we plot the exact one-loop expression
for $\Delta_{SQCD}$ and the expansions of eqs.\ \ref{eq:0degexpansion}
and \ref{eq:heavyb2}, as a function of $M_{\tilde b_2}$, for
$M_{\tilde b_1} = M_{\tilde g} = \mu = A_b = M_A = 200$ GeV
and three different values of $\tan\beta$.
Clearly, in order for $\Delta_{\rm SQCD}$ to be large in
the case of near-zero sbottom mixing, both of the sbottoms must
be reasonably light.  Note however that, due to the enhancement in
$\tan\beta$, the $1/M_{\tilde b_2}^2$ suppression is not so small.
As an example, for $\tan\beta=50$,
$M_{\tilde b_1} = M_{\tilde g} = \mu = A_b = M_A = 200$ GeV
and $M_{\tilde b_2}=500$~GeV
[1000~GeV], one obtains $\Delta_{\rm SQCD} \simeq -10\%$ [$-5\%$].

\begin{figure}[t]
\begin{center}
\resizebox{10cm}{!}{\rotatebox{270}{\includegraphics{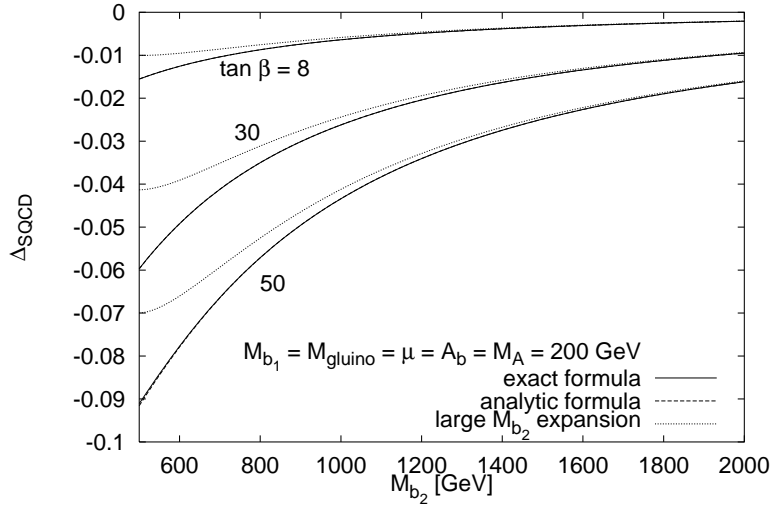}}}
\caption{$\Delta_{\rm SQCD}$ as a function of $M_{\tilde b_2}$,
with $M_{\tilde b_1} = M_{\tilde g} = \mu = A_b = M_A = 200$ GeV and
$\tan\beta = 8, 30, 50$.  Solid lines are based on the
exact one-loop formula, dashed lines based on are the analytic expansion
for near-zero sbottom mixing of eq.\ \ref{eq:0degexpansion},
and dotted lines are based on the expansion for large $M_{\tilde b_2}$
of eq.\ \ref{eq:heavyb2}.}
\label{fig:0deg:1aplot}
\end{center}
\end{figure}

The various cases examined in this section can be summarized by
specifying the behavior of $C_1$ and $C_2$ of eq.\ \ref{genform} on the
model parameters.  In Table~1, four cases are shown.  In all cases,
$M_{SUSY}$
is identified with the largest supersymmetry-breaking mass, while $M$
refers to a possible intermediate supersymmetric mass scale (with
$M_{EW}\ll M\ll M_{SUSY}$).  The presence of a factor of $\tan\beta$
(unless multiplied by $M/M_{SUSY}$)
indicates delayed decoupling.  In the case of $M_{\tilde g}=M_{SUSY}$,
$C_2\sim (M_{SUSY}/M)\tan\beta$ implies a delayed decoupling that
vanishes only as a single power of $1/M_{SUSY}$.  Finally, in the case
of large $M_{\tilde b_2}$, $C_2\sim M^2_{SUSY}/M^2$ implies no decoupling
as $M_{SUSY}\to\infty$ with $M$ held fixed.  This is not a violation
of the usual decoupling theorem \cite{dhp,ac}, since in the latter case,
only part of the supersymmetric spectrum has been removed from the
low-energy effective theory.  Decoupling is recovered in the limit
of $M\to\infty$, as expected.

\vspace{0.2in}
\begin{table}[htbp]
\begin{center}
\begin{tabular}{|c|c| c c |}\hline
Case & $\widetilde b$ mixing & $C_1$ & $C_2$  \rule{0in}{3ex} \\[1ex] \hline
 $M_S\simeq M_{\tilde g}=M_{SUSY}$ & maximal & $\tan\beta$ & $\tan\beta$
\rule{0in}{3ex} \\
 $M_S=M_{SUSY}\gg M$ & maximal & $(M^2/ M_{SUSY}^2)\tan\beta$ & 1
 \rule{0in}{3ex} \\
 $M_{\tilde g}=M_{SUSY}\gg M$ & maximal & $(M/M_{SUSY})\tan\beta$
& $(M_{SUSY}/ M)\tan\beta$ \rule{0in}{3ex} \\
 $M_{\tilde b_2}=M_{SUSY}\gg M$ & near-zero & $(M^2/
M^2_{SUSY})\tan\beta$ & $M^2_{SUSY}/M^2$ \rule{0in}{3ex} \\
\hline
\end{tabular}
\vspace{.2in}
\parbox{5in}{\caption [0]
{\label{sumtable} Approach to decoupling of the one-loop
${\mathcal O}(\alpha_s)$ radiative
corrections to the $h^0b\bar b$ vertex: $\Delta_{SQCD}\sim C_1
(M_{EW}^2/M_A^2) + C_2(M_{EW}^2/M_{SUSY}^2)$.  See text for further
discussion.}}
\vspace{-0.3in}
\end{center}
\end{table}

\newpage
\section{Conclusions}
\label{sec:conclusions}
In this paper we have studied the one loop SQCD corrections to the
$h^0 b \bar b$ coupling in the limit of large SUSY masses.  In order
to understand analytically the behavior of the corrections in this
limit,
we have performed expansions for the SUSY mass parameters large compared
to the electroweak scale.
We have shown that for the SUSY mass parameters and $M_A$ large and all of the
same order, the SQCD corrections decouple like the inverse
square of these mass parameters.  However, if the mass parameters are not
all of the same size, then this behavior can be modified.  If $M_A$
is light, then the SQCD corrections to the $h^0 b \bar b$ coupling
generically do not decouple in the limit of large SUSY mass parameters.
In this case, the low-energy theory at the electroweak scale contains
two full Higgs doublets with Higgs-fermion couplings of the general
type-III model.

We have also examined three cases in which there is a hierarchy among
the SUSY mass parameters.  In the case of maximal sbottom mixing with
$M_S$ large and the other SUSY mass parameters and $M_A$ of order a
common mass scale
$M$ (chosen such that $M_{EW}\ll M\ll M_S$), the SQCD
corrections decouple like $M^2/M_S^2$.
Second, we examined the case of a large gluino mass with the other SUSY
mass parameters of order a common mass scale $M$
(chosen such that $M_{EW}\ll M\ll M_{\tilde g}$).
In this case we found that the SQCD corrections
decouple more slowly, like
$(M/M_{\tilde g}) \log(M^2_{\tilde g}/M_S^2)$.
Finally, we examined the case in which one sbottom is much heavier than
the other SUSY mass parameters, which are fixed at a scale $M$.
In this case the mixing angle in the sbottom sector is near zero.
We found that the
piece of the SQCD corrections that is enhanced at large $\tan\beta$
decouples
like $M^2/M_{\tilde b_2}^2$.  There is also a piece of the SQCD
corrections that does not decouple as $M_{\tilde b_2}$ is taken large,
but it is not enhanced by $\tan\beta$
and is numerically negligible compared to the decoupling piece, 
up to a very high value of the heavier
sbottom mass.

The decoupling behavior of the SQCD corrections to the $h^0 b \bar b$
coupling implies that distinguishing the lightest MSSM Higgs boson
from the SM Higgs boson will be very difficult if $A^0$
and the SUSY spectrum are heavy, even after one-loop SUSY corrections
are taken into account.  However, because
of the enhancement at large $\tan\beta$, the onset of decoupling is
delayed, and the corrections can still
be at the percent level for $\tan\beta \sim 50$ and all SUSY mass
parameters and $M_A$ of order 1 TeV.  If one or both of the sbottoms,
the gluino,
and/or $A^0$ lie below the TeV scale, then the SQCD corrections will be
larger still.  The decoupling limit provides a challenge for Higgs
searches at future colliders.  Even if the light CP-even Higgs boson is
found, the direct discovery of supersymmetric particles may be essential for
unraveling the origin of electroweak symmetry breaking.

\newpage
\begin{center}
{\Large \bf Acknowledgments}
\end{center}

We wish to thank M.~Carena,  K.~Matchev, C.~E.~M.~Wagner and
G.~Weiglein for interesting discussions. M.H., S.P., S.R. and
D.T. kindly thank T.~Hahn for providing the code of LoopTools
(which was used in the calculation) and for many helpful
suggestions. H.L. is grateful to G.~Kribs for a discussion on
models of SUSY breaking.

This work has been supported in part by
the Spanish Ministerio de Educacion y Cultura under project CICYT
AEN97-1678.  S.R. has been partially supported by the European
Union through contract ERBFMBICT972474. H.E.H. is supported in part
by the U.S. Department of Energy under contract DE-FG03-92ER40689.
Fermilab is operated by Universities Research Association Inc.\
under contract no.~DE-AC02-76CH03000 with the U.S. Department of
Energy.

\begin{center}
{\Large \bf Appendix}
\end{center}

\appendix
\section{Expansions of loop functions}
\label{sec:loopintexpansions}
In this Appendix we define our notation for the two- and three-point
integrals that appear in eqs.\ \ref{eqn:fulldgvertex}
and \ref{eqn:fulldgwfct}
and give formulae for their expansions in powers of the SUSY mass
parameters.

We follow the definitions and conventions of \cite{Hollik}.
The two-point integrals are given by:
\begin{equation}
    \mu^{4-D} \int \frac{d^D k}{(2\pi)^D} \frac{\{1; k^{\mu}\}}
    {[k^2 - m_1^2][(k+q)^2 - m_2^2]}
    = \frac{i}{16\pi^2} \left\{ B_0; q^{\mu} B_1 \right\}
    (q^2;m_1^2,m_2^2)\,.
\end{equation}
The derivatives of the two-point functions are defined as follows:
\begin{equation}
    B^{\prime}_{0,1}(p^2;m_1^2,m_2^2)
    = \left. \frac{\partial}{\partial q^2} B_{0,1}(q^2;m_1^2,m_2^2)
    \right|_{q^2 = p^2}\,.
\end{equation}
Finally, the three-point integrals are given by:
\begin{Eqnarray}
    &\mu^{4-D}& \int \frac{d^D k}{(2\pi)^D}
    \frac{\{1; k^{\mu}\}}
    {[k^2 - m_1^2][(k + p_1)^2 - m_2^2][(k + p_1 + p_2)^2 - m_3^2]}
    \nonumber \\
    & & = \frac{i}{16\pi^2}
    \left\{ C_0; p_1^{\mu}C_{11} + p_2^{\mu}C_{12} \right\}
    (p_1^2, p_2^2, p^2; m_1^2, m_2^2, m_3^2)\,,
\end{Eqnarray}

\vspace{-0.1in}\noindent
where $p = -p_1 - p_2$.

We now give the large $M_{SUSY}$ expansions of the loop integrals.

\subsection{Maximal $\tilde b_L - \tilde b_R$ mixing}
The loop integrals are expanded as follows, where $M_S^2$ and $\Delta^2$
are defined in eq.\ \ref{eq:MSandDelta}.
\begin{Eqnarray}
    &&\hspace{-1cm} C_0(m_b^2,M_{h^0}^2,m_b^2;M_{\tilde g}^2, M_{\tilde b_1}^2,
    M_{\tilde b_1}^2)
    \nonumber \\
    &&\qquad \simeq -\frac{1}{2 M_S^2} f_1(R)
    + \frac{\Delta^2}{3 M_S^4} f_2(R)
    - \frac{\Delta^4}{4 M_S^6} f_3(R)
    - \frac{M_{h^0}^2}{24 M_S^4} f_4(R)  \nn \\
    &&\hspace{-1cm} C_0(m_b^2,M_{h^0}^2,m_b^2;M_{\tilde g}^2, M_{\tilde b_2}^2,
    M_{\tilde b_2}^2)
    \nonumber \\
    &&\qquad \simeq -\frac{1}{2 M_S^2} f_1(R)
    - \frac{\Delta^2}{3 M_S^4} f_2(R)
    - \frac{\Delta^4}{4 M_S^6} f_3(R)
    - \frac{M_{h^0}^2}{24 M_S^4} f_4(R) \nn \\
    &&\hspace{-1cm} C_0(m_b^2,M_{h^0}^2,m_b^2;M_{\tilde g}^2,M_{\tilde b_1}^2,
    M_{\tilde b_2}^2)
	\nonumber \\
    &&\qquad \simeq -\frac{1}{2 M_S^2} f_1(R) 
	- \frac{\Delta^4}{12 M_S^6} f_3(R)
	- \frac{M_{h^0}^2}{24 M_S^4} f_4(R)
	\nn \\
    &&\hspace{-1cm} C_{11}(m_b^2,M_{h^0}^2,m_b^2;M_{\tilde g}^2, M_{\tilde b_1}^2,
    M_{\tilde b_1}^2)
	\nonumber \\
    &&\qquad \simeq \frac{1}{3 M_S^2} f_5(R)
    - \frac{\Delta^2}{4 M_S^4} f_4(R) 
	+ \frac{\Delta^4}{5 M_S^6} f_6(R)
	+ \frac{M_{h^0}^2}{30 M_S^4} f_7(R)
	\nn \\
    &&\hspace{-1cm} 
      C_{11}(m_b^2,M_{h^0}^2,m_b^2;M_{\tilde g}^2, M_{\tilde b_2}^2,
    M_{\tilde b_2}^2)
	\nn \\
    &&\qquad \simeq \frac{1}{3 M_S^2} f_5(R)
    + \frac{\Delta^2}{4 M_S^4} f_4(R) 
	+ \frac{\Delta^4}{5 M_S^6} f_6(R)
	+ \frac{M_{h^0}^2}{30 M_S^4} f_7(R)
	\nn \\
    &&\hspace{-1cm} B_0(m_b^2;M_{\tilde g}^2,M_{\tilde b_1}^2)
    - B_0(m_b^2;M_{\tilde g}^2,M_{\tilde b_2}^2)
    \simeq -\frac{\Delta^2}{M_S^2} f_1(R) 
	\nn \\
    &&\hspace{-1cm} B_0^{\prime}(m_b^2;M_{\tilde g}^2,M_{\tilde b_1}^2)
    - B_0^{\prime}(m_b^2;M_{\tilde g}^2,M_{\tilde b_2}^2)
    \simeq 
    -\frac{\Delta^2}{6 M_S^4} f_8(R) \nn \\
    &&\hspace{-1cm} B_1^{\prime}(m_b^2;M_{\tilde g}^2,M_{\tilde b_1}^2)
    + B_1^{\prime}(m_b^2;M_{\tilde g}^2,M_{\tilde b_2}^2)
    \simeq  -\frac{1}{6 M_S^2} f_4(R)
	- \frac{\Delta^4}{15 M_S^6} f_9(R).
\end{Eqnarray}

\vspace{-0.1in}\noindent
The functions $f_i(R)$ are given in terms of the ratio
$R \equiv M_{\tilde g}/M_S$:
\begin{Eqnarray} \label{thefi}
    f_1(R) &=& \frac{2}{(1-R^2)^2} \left[1 - R^2 + R^2 \log R^2\right]
        \nonumber \\
    f_2(R) &=& \frac{3}{(1-R^2)^3} \left[1 - R^4 + 2 R^2 \log R^2\right]
        \nonumber \\
    f_3(R) &=& \frac{4}{(1-R^2)^4} \left[1 + \nicefrac{3}{2}R^2 - 3R^4
        + \nicefrac{1}{2}R^6 + 3 R^2 \log R^2\right] \nonumber \\
    f_4(R) &=& \frac{4}{(1-R^2)^4} \left[\nicefrac{1}{2} - 3R^2
        + \nicefrac{3}{2} R^4 + R^6 - 3 R^4 \log R^2\right]
        \nonumber \\
    f_5(R) &=& \frac{3}{(1-R^2)^3} \left[ \nicefrac{1}{2} - 2R^2
        + \nicefrac{3}{2}R^4 - R^4 \log R^2 \right]
    \nonumber \\
	f_6(R) &=& \frac{5}{(1-R^2)^5} \left[ \nicefrac{1}{2} - 4R^2
		+ 4R^6 - \nicefrac{1}{2}R^8 - 6R^4 \log R^2 \right]
		\nonumber \\
	f_7(R) &=& \frac{5}{(1-R^2)^5} \left[ \nicefrac{1}{3} - 2R^2
		+ 6R^4 - \nicefrac{10}{3}R^6 - R^8
		+ 4R^6 \log R^2 \right] \nonumber \\
    f_8(R) &=& \frac{12}{(1-R^2)^4} \left[ \nicefrac{1}{2} + 2R^2
       - \nicefrac{5}{2} R^4 + 2R^2 \log R^2 + R^4 \log R^2
       \right] \nn \\
    f_9(R) &=& \frac{5}{(1-R^2)^6} \left[ 1 - 12R^2 - 36R^4
	+ 44R^6 + 3R^8
	\right. \nn \\ 
   &&\qquad\qquad\qquad\qquad 
   \left. - 24R^6 \log R^2 - 36R^4 \log R^2 \right]\,.
\end{Eqnarray}

\vspace{-0.1in}\noindent
Note that in the special case $M_{\tilde g} = M_S$, $R=1$ and
$f_i(1)=1$.

\subsection{Near-zero $\tilde b_L - \tilde b_R$ mixing}
The loop integrals are expanded as follows:
\begin{Eqnarray}
 &&\hspace{-1.3cm} C_0(m_b^2,M_{h^0}^2,m_b^2;M_{\tilde g}^2, M_{\tilde b_1}^2,
    M_{\tilde b_1}^2)
    \simeq -\frac{1}{2M_{\tilde b_1}^2} f_1(R_1)  
	- \frac{M_{h^0}^2}{24 M_{\tilde b_1}^4} f_4(R_1)\nn \\
 &&\hspace{-1.3cm} C_0(m_b^2,M_{h^0}^2,m_b^2;M_{\tilde g}^2, M_{\tilde b_2}^2,
    M_{\tilde b_2}^2)
    \simeq -\frac{1}{2M_{\tilde b_2}^2} f_1(R_2) 
	- \frac{M_{h^0}^2}{24 M_{\tilde b_2}^4} f_4(R_2)\nn \\
    &&\hspace{-1.3cm} 
      C_0(m_b^2,M_{h^0}^2,m_b^2;M_{\tilde g}^2, M_{\tilde b_1}^2,
    M_{\tilde b_2}^2)
    \simeq
    -\frac{h_1(R_1,R_2)}{(M_{\tilde b_1}^2 - M_{\tilde b_2}^2)}
    + \frac{M_{h^0}^2 \, h_2(R_1,R_2)}
    {(M_{\tilde b_1}^2 - M_{\tilde b_2}^2)^2}
    \nn \\
    &&\hspace{-1.3cm}
      C_{11}(m_b^2,M_{h^0}^2,m_b^2;M_{\tilde g}^2, M_{\tilde b_1}^2,
    M_{\tilde b_1}^2)
    \simeq \frac{1}{3M_{\tilde b_1}^2} f_5(R_1)  
	+ \frac{M_{h^0}^2}{30 M_{\tilde b_1}^4} f_7(R_1)\nn \\
    && \hspace{-1.3cm} 
    C_{11}(m_b^2,M_{h^0}^2,m_b^2;M_{\tilde g}^2, M_{\tilde b_2}^2,
    M_{\tilde b_2}^2)
    \simeq \frac{1}{3M_{\tilde b_2}^2} f_5(R_2)  
	+ \frac{M_{h^0}^2}{30 M_{\tilde b_2}^4} f_7(R_1)\nn \\
    && \hspace{-1.3cm} B_0(m_b^2; M_{\tilde g}^2, M_{\tilde b_1}^2)
    - B_0(m_b^2; M_{\tilde g}^2, M_{\tilde b_2}^2)
    \simeq - h_1(R_1,R_2) 
	\nn \\
    && \hspace{-1.3cm} B_0^{\prime}(m_b^2; M_{\tilde g}^2, M_{\tilde b_1}^2)
    - B_0^{\prime}(m_b^2; M_{\tilde g}^2, M_{\tilde b_2}^2)
    \simeq 
    \frac{1}{6M_{\tilde b_1}^2} f_2(R_1)
    - \frac{1}{6M_{\tilde b_2}^2} f_2(R_2) \nn \\
    &&\hspace{-1.3cm} B_1^{\prime}(m_b^2; M_{\tilde g}^2, M_{\tilde b_1}^2)
    + B_1^{\prime}(m_b^2; M_{\tilde g}^2, M_{\tilde b_2}^2)
    \simeq -\frac{1}{12 M_{\tilde b_1}^2} f_4(R_1)
    - \frac{1}{12 M_{\tilde b_2}^2} f_4(R_2),
\end{Eqnarray}

\vspace{-0.1in}\noindent
where $R_i \equiv M_{\tilde g}/M_{\tilde b_i}$ ($i=1,2$).
The functions $f_i(R)$ were given in eq.\ \ref{thefi}.
The functions $h_1(R_1,R_2)$ and $h_2(R_1,R_2)$ are defined as follows:
\begin{Eqnarray}
    h_1(R_1,R_2) &=& h(R_1) - h(R_2), \qquad {\rm with} \qquad
    h(R) = -\frac{\log R^2}{1-R^2},  \nn \\
    h_2(R_1,R_2) &=& 1 +
    \frac{R_1^2 + R_2^2 - 2 R_1^2 R_2^2}{2 (1-R_1^2) (1-R_2^2)}
    \nn \\[5pt]
    &-& \frac{1}{2 (R_1^2 - R_2^2)} \left[
    \frac{\log R_1^2}{(1-R_1^2)^2} (R_1^2 + R_2^2 - 2 R_1^4) \right.
    \nn \\[5pt]
    & & \left. \qquad\qquad\qquad - \frac{\log R_2^2}{(1-R_2^2)^2}
    (R_1^2 + R_2^2 - 2 R_2^4) \right]\,.
\end{Eqnarray}

\vspace{-0.1in}\noindent
The functions $h$ and $h_2$ have the following
properties:
\begin{Eqnarray}
    h(1) &=& 1, \nn \\[5pt]
    h_2(R_1,R_2) &=& h_2(R_2,R_1), \nn \\[5pt]
    h_2(1,R_2) &=& \frac{1}{(1-R_2^2)^2}\left[
    \nicefrac{5}{4} - R_2^2 - \nicefrac{1}{4} R_2^4 +
    \left(\nicefrac{1}{2} + R_2^2\right) \log R_2^2 \right]\,.
\end{Eqnarray}


\end{document}